\documentclass[12pt]{article}

\usepackage{amssymb}
\usepackage{amsmath}
\usepackage{graphicx}
\usepackage{latexsym}

\begin{document}

\begin{titlepage}
\begin{center}

\hfill UT-HET 024 \\
\hfill STUPP-09-200 \\
\hfill \today

\vspace{0.5cm}
{\large\bf Productions of second Kaluza-Klein gauge bosons \\ 
in the minimal universal extra dimension model at LHC}
\vspace{1cm}

{\bf Shigeki Matsumoto}$^{a,\,}
$\footnote{smatsu@sci.u-toyama.ac.jp},
{\bf Joe Sato}$^{b,\,}
$\footnote{joe@phy.saitama-u.ac.jp},
{\bf Masato Senami}$^{c,\,}
$\footnote{senami@me.kyoto-u.ac.jp}, \\
and 
{\bf Masato Yamanaka}$^{b,\,}
$\footnote{masa@krishna.phy.saitama-u.ac.jp} \\

\vskip 0.15in

{\it
$^a${Department of Physics, University of Toyama, Toyama 930-8555, Japan} \\
$^b${Department of Physics, Saitama University, 
     Shimo-okubo, Sakura-ku, \\Saitama, 338-8570, Japan} \\
$^c${Department of Micro Engineering,
      Kyoto University, Kyoto 606-8501, Japan \\
      This author was in ICRR, Univ. of Tokyo, when this work started. }
}

\vskip 0.4in

%%%%%%%%%%%%%%%%%%%%%%%%%%%%%%%%%%%%%%%%%%%%%%%%%%%%%%%%%%%%%%%%%%
\abstract{We calculate the production rates of the second Kaluza-Klein (KK) 
photon $\gamma^{(2)}$ and  Z boson $Z^{(2)}$ at the LHC including all 
significant processes in the minimal Universal Extra Dimension (MUED) model. 
%%%
For discrimination of the MUED model from other TeV scale models at the LHC, 
$\gamma^{(2)}$ and $Z^{(2)}$ play a crucial role. In order to discuss the 
discrimination and calculate their production rates, we derive KK number 
violating operators including the contribution of the top Yukawa coupling. 
%%%
Using these operators,  we accurately calculate branching 
ratios of second KK particles.
%%%
In addition we find that these KK number violating operators provide  
new processes for $\gamma^{(2)}$ and $Z^{(2)}$ productions, such as 
cascade decay from second KK quarks produced through these operators.
They have large contributions to their total production rates. In particular, 
these production processes give the dominant contribution for $\gamma^{(2)}$ 
production for $1/R \gtrsim 800$ GeV. 
%%%
As a result, with an integrated luminosity of 100 fb$^{-1}$, the number of 
produced $\gamma^{(2)}$ and $Z^{(2)}$ are estimated as 10$^6$ - 10$^2$ 
for the compactification scale between 400 GeV and 2000 GeV. }
%%%%%%%%%%%%%%%%%%%%%%%%%%%%%%%%%%%%%%%%%%%%%%%%%%%%%%%%%%%%%%%%%

\end{center}
\end{titlepage}
\setcounter{footnote}{0}

%%%%%%%%%%%%%%%%%%%%%%%%%%%%%%%%%%%%%%%%%%%%%%%%%%%%%%%%%%%%%%%%
\section{Introduction} %%%%%%%%%%%%%%%%%%%%%%%%%%%%%%%%%%%%%%%%%%%%%%%%%%%%%
%%%%%%%%%%%%%%%%%%%%%%%%%%%%%%%%%%%%%%%%%%%%%%%%%%%%%%%%%%%%%%%%

Theoretical arguments in particle physics and cosmological observations
conclude that the Standard Model (SM) is not the theory of everything
but is rather the effective theory describing physics below ${\cal O} (100)$
GeV. Many models beyond the SM have been proposed, and the Universal Extra
Dimension (UED) models \cite{Appelquist:2000nn} are some of the attractive
candidates for new physics at TeV scale\footnote{UED models are motivated 
by the TeV scale extra dimension theory \cite{Antoniadis}.}.
In the UED models, all SM fields can propagate into compactified extra
dimensions, and hence they are accompanied by the tower of Kaluza-Klein
(KK) particles. They can give plausible explanations for the existence of 
dark matter \cite{Cheng:2002ej}, the number of
fermion generations \cite{Dobrescu:2001ae}, SM neutrino masses which are 
embedded in extended models \cite{Matsumoto:2006bf}, and so on.
Among various UED models, the simplest and the most popular one is
called the Minimal UED (MUED) model. The MUED model is defined on the
five dimensional space-time, where the extra dimension is compactified
on an $S^1/Z_2$ orbifold. In the MUED model, only two parameters are
newly introduced to the SM. One is the compactification scale of the
extra dimension $1/R$, the inverse of the radius of $S^1$ circle, and
the other is the cutoff scale of the MUED model $\Lambda$. 
In order to satisfy terrestrial experiments, $1/R$ must be larger than
400 GeV \cite{Appelquist:2000nn, Agashe:2001ra}, while the calculation of the
relic abundance of the lightest KK particle (LKP) suggests that the 
abundance of the dark matter is explained for 500
GeV $\lesssim 1/R \lesssim$ 1500 GeV \cite{Kakizaki:2005en}. As
indicated from these results, the MUED model would be discovered and
studied at the Large Hadron Collider (LHC).
%Due to the $S^1/Z_2$ orbifold, the KK particles are always produced in 
%a pair, and hence the signals of the extra dimension appear only through the loop suppressions\footnote{To be exact, 
%as shown in the section 4, there is a case that the signals of the extra dimension can be observed at tree level. 
%However, those signals are kinematically suppressed.}. Therefore phenomenological constraints on the MUED model
%is not so stringent. As a result of many studies \cite{Appelquist:2002wb, Gogoladze}, the electroweak 
%precision measurements constrain the compactification scale as $1/R \gtrsim$ 400 GeV. Here $R$ is a radius of
%$S^1$ manifold.

The confirmation of the MUED model (in general, UED models) at collider
experiments requires the discovery of KK particles. Though the LHC 
can produce KK particles, it is difficult to
confirm that they are indeed the KK particles, because new particles
predicted in various TeV scale models give quite similar signatures to
each other. Therefore it is very important to understand how they can be
identified as KK particles.
In this article, we discuss the discrimination of the MUED model from
other TeV scale models.

In Ref. \cite{Datta:2005zs}, an excellent idea to discriminate the
MUED model from other TeV scale models has been proposed. The essence of the
discrimination is the discovery of the second KK particles.  The signals
of the first KK particles are quite similar to those of new
particles in other models.
However, the discovery of the second KK particles strongly suggests the
existence of the MUED model, since their masses are peculiarly almost 
equal to $2/R$, and the value of $1/R$ is expected by the masses of 
the ``first KK'' particles.
In particular, the second KK photon
$\gamma^{(2)}$ and Z boson $Z^{(2)}$ play an important
role for the search of the second KK particles. They are able to decay
into two charged leptons. It is possible to reconstruct the masses
of $\gamma^{(2)}$ and $Z^{(2)}$ clearly from the charged dileptons
emitted by them. Connecting their masses and those of the first KK
particles, we can confirm the realization of the MUED model.

Hence, $\gamma^{(2)}$ and $Z^{(2)}$ are the key ingredients to discriminate
the MUED model from other TeV scale models. We therefore calculate
their production rates at the LHC. To do this, first we 
derive the effective Lagrangian containing KK number violating operators, 
which are relevant to both the second KK particles productions and decay. Previously, 
these operators have been discussed in Ref. \cite{Cheng:2002iz} considering only 
gauge interactions. We improve the operators by including the contribution 
of Yukawa interactions.
%%%
Through the KK number violation operators, $\gamma^{(2)}$ and
$Z^{(2)}$ decay into dilepton. In addition to the decay, these operators allow 
single second KK particle production, which does not suffer from a severe 
kinematical suppression compared to pair productions. 
Next we study the production processes of $\gamma^{(2)}$ and
$Z^{(2)}$ including both the KK number violating 
and conserving processes. At the LHC, $\gamma^{(2)}$ and $Z^{(2)}$
are produced mainly through the cascade decays of the second KK gluons
$g^{(2)}$ and quarks $q^{(2)}$. Here the symbol $q$ stands 
for both $SU(2)$ doublet and singlet quarks. We also
need to calculate their branching ratios into $\gamma^{(2)}$ and $Z^{(2)}$.
Finally, we calculate the production rates of $\gamma^{(2)}$ and $Z^{(2)}$
from each process, and estimate the number of the dilepton signals from them.

This article is organized as follows. In the next section, we briefly review the 
MUED model. Then we mention the difficulty that appears during the confirmation of 
the MUED model at the LHC, and discuss an idea to overcome the difficulty. 
In Sec. 3, we discuss the Lagrangian relevant to the productions of $\gamma^{(2)}$ 
and $Z^{(2)}$, and we calculate the branching ratios of colored second KK particles. 
In Sec. 4, we show some numerical results for 
$\gamma^{(2)}$ and $Z^{(2)}$ productions, and discuss the significance to discover 
the particles at the LHC. Section 5 is devoted to summary.

\vspace{0mm}
%%%%%%%%%%%%%%%%%%%%%%%%%%%%%%%%%%%%%%%%%%%%%%%%%%%%%%%%%%%%%%
\section{The MUED model and its discrimination from other models}  %%%%%%%%%%%%%%%%%%%%%%%%%%%%%%
%%%%%%%%%%%%%%%%%%%%%%%%%%%%%%%%%%%%%%%%%%%%%%%%%%%%%%%%%%%%%%

The MUED model is the simplest version of the UED models. In this model, fields 
with a fifth dimensional momentum $n/R$ behave as new heavy particles with 
a tree-level mass $\sqrt{m_{SM}^2 + (n/R)^2}$ from the viewpoint of four 
dimensional field theory. These new particles are called KK particles, $n$ is 
called the KK number ($n$ = 0 for SM particles, $n$ = 1, 2, ... for KK particles), 
and $m_{SM}$ represents the mass of the corresponding SM particle.

The SM particles and their KK
particles have identical gauge charges and spins. All
interactions of the KK particles in the four dimensional space-time are
determined by the Lagrangian in five dimensions. Since the UED models are not
renormalizable, they should be considered as an effective theory
defined at the scale $\Lambda $, where $\Lambda$ is usually taken to be
$\Lambda R \sim {\cal O} (10)$ \cite{Appelquist:2000nn,
Bhattacharyya:2006ym}. In this article, we take $\Lambda$ to be $\Lambda
R = 20$\footnote{Our results are almost independent of $\Lambda$, since
it always appears with a loop suppression and gives only logarithmic
corrections.}. Thus, in order to discuss the phenomenology of the
MUED model, we need only two new parameters : 
$1/R$ and $\Lambda$.

Although the masses of KK particles at tree-level are highly degenerate in each 
KK mode, the degeneracy is slightly relaxed by radiative corrections 
\cite{Cheng:2002iz}. The mass spectrum of the $n$-th KK particles with the 
radiative corrections $\delta m_n$ are given by 
$m_n = \sqrt{ (n/R)^2 + m_{SM}^2 + \delta m_n^2 }$.  Here, analytical 
expressions of the radiative corrections $\delta m_n$ are given in Ref. 
\cite{Cheng:2002iz}. In general, due to the radiative corrections, colored KK 
particles are heavier than non-colored KK particles in each KK level.  
%%%
We have used couplings improved by the renormalization group (RG) equation 
to compute the radiative corrections to the masses of KK particles. 
The gauge couplings at the one-loop level are given by
\begin{equation}
 \begin{split}
   \alpha_i^{-1} (\mu) = \alpha_i^{-1} (m_Z) - \frac{b_i}{2 \pi} \text{ln} \bigl[ \mu/m_Z \bigr] 
   - \sum_n \theta(\mu - n/R) \frac{\tilde b_i}{2 \pi} \text{ln} \bigl[ \mu/(n/R) \bigr] ~.
 \end{split}    \label{RG}
\end{equation} 
Here $\alpha_i = g_i^2/(4 \pi)$, $g_i = (g', g_2, g_s)$ are the SM gauge 
coupling constants ($i$ stands for each gauge group), $\mu$ is the renormarization 
scale, and $\theta$ represents the step function. $b_i$ and $\tilde b_i$ are 
summarized in Table 1.
\begin{table}[!ht]
\begin{center}
\begin{tabular}{ | c || c | c | c | c | c | c |} \hline 
 Fields & $b'$ & $b_2$ & ~$b_s$~ & $\tilde b'$ & $\tilde b_2$ & $\tilde b_s$  \\ \hline
 Gauge & 0 & $-22/3$ & $-11$ & 0 & $-7$ & $-21/2$     \\ \hline
 Higgs & 1/6 & 1/6 & 0 & 1/6 & 1/6 & 0      \\ \hline 
 Fermion & 20/3 & 4 & 4 & 40/3 & 8 & 8    \\ \hline 
\end{tabular}  \\[2mm]
\caption{{\small{Coefficients of RG improved gauge coupling constants. Each $\tilde b_i$ is the contribution 
                             from one KK level of KK particles.}}  }
\end{center}      \label{RG table}
\end{table}   \\
%%%
The mass spectrum of the first KK particles ($n=1$) and the second KK particles 
($n=2$) for $1/R = 500$ GeV are shown in Table 2. 
%%%
When we calculate the mass spectrum of n-th KK particles, we choose their mass 
scale, $n/R$, as renormalization scale.
%%%
It is seen that each KK mode is degenerate with each other and the masses of 
the second KK particles are almost twice of the first KK particles.

\begin{table}[!ht]
\begin{center}
\begin{tabular}{ | l || l | l |} \hline 
~~~~~~~~~~~~~~KK particle                         & ~~~~~~$n=1$~~~~~~ & ~~~~~~$n=2$~~~~~~~    \\ \hline
 KK gluon ~$g^{(n)}$                                &    ~~~~618 GeV      &  ~~~~1170 GeV         \\ \hline
 KK Z boson ~$Z^{(n)}$                             &   ~~~~534 GeV       &  ~~~~1059 GeV         \\ \hline 
 KK W boson ~$W^{\pm (n)}$                     &   ~~~~534 GeV       &  ~~~~1046 GeV         \\ \hline 
 KK photon ~$\gamma^{(n)}$                      &   ~~~~501 GeV       &  ~~~~1000 GeV         \\ \hline
 KK Higgs ~$H^{(n)}$                                &   ~~~~518 GeV       &   ~~~~1014 GeV        \\ \hline
 KK CP-odd Higgs ~$A^{(n)}$                      &   ~~~~512 GeV       &   ~~~~1011 GeV        \\ \hline
 KK charged Higgs ~$H^{\pm (n)}$               &   ~~~~510 GeV       &   ~~~~1010 GeV             \\ \hline 
 KK $SU(2)$ singlet electron ~$E^{(n)}$       &  ~~~~505 GeV       &   ~~~~1008 GeV             \\ \hline 
 KK $SU(2)$ doublet lepton ~$L^{(n)}$         &  ~~~~514 GeV       &   ~~~~1022 GeV             \\ \hline 
 KK $SU(2)$ singlet up quark ~$U^{(n)}$       &  ~~~~571 GeV       &   ~~~~1100 GeV             \\ \hline 
 KK $SU(2)$ singlet down quark ~$D^{(n)}$   &  ~~~~569 GeV       &   ~~~~1097 GeV             \\ \hline 
 KK $SU(2)$ doublet quark ~$Q^{(n)}$          &  ~~~~582 GeV       &   ~~~~1117 GeV              \\ \hline
 KK light top quark ~$T^{(n)}$                     &  ~~~~569 GeV       &    ~~~~1071 GeV            \\ \hline 
 KK heavy top quark ~$t^{(n)}$                    &  ~~~~594 GeV       &    ~~~~1109 GeV            \\ \hline
 KK $SU(2)$ singlet bottom quark ~$B^{(n)}$ &  ~~~~569 GeV       &    ~~~~1097 GeV            \\ \hline 
 KK $SU(2)$ doublet bottom quark ~$b^{(n)}$ &  ~~~~575 GeV       &    ~~~~1106 GeV            \\ \hline 
\end{tabular}  \\[2mm]
\caption{{\small{Mass spectrum of KK particles for $1/R$ = 500 GeV and $\Lambda R$ = 20. }}  }
\end{center}
\end{table}

Since the translational invariance along the extra dimension direction
is broken due to the orbifolding on an $S^1/Z_2$, the fifth
dimensional momentum (KK number) is no longer conserved. Nevertheless, the
subgroup of the translational invariance remains unbroken, which is called the
KK parity. Under the parity, particles with even (odd) KK number have
plus (minus) sign, and the product of the sign is conserved in each
process. Because of the KK parity conservation, the lightest KK particle
(LKP) is stable and provided as a candidate for dark matter. The situation
is quite similar to the case of supersymmetric models with R-parity
conservation, in which the Lightest Supersymmetric Particle (LSP) is
stable. The relic abundance of the KK particle dark matter has been
calculated in the MUED model \cite{Kakizaki:2005en},
and it turns out that if the KK particle dark matter dominates the
component of dark matter in the universe, $1/R$ should be in the range
of 500 GeV-1500 GeV. The MUED model, therefore,  would be explored at the
LHC.

The confirmation of the MUED model at the LHC is not so easy.  Although it is
necessary to discover the KK particles, it is very hard to
distinguish the signals of these particles from those of other TeV scale
models (for example, the MSSM, the little Higgs models, and so on), because
these models also predict new heavy particles which give quite similar
signals.
For example, in a supersymmetric model, new particles also have identical couplings 
to their corresponding SM particles and the masses of colored new particles are 
heavier than those of other new particles. Furthermore, there is a conserved discrete 
symmetry (R parity), which makes LSP stable. In both models, at the LHC, heavy 
colored new particles are produced first. Then they immediately decay into lighter 
new particles. Finally the lightest new particles leave detectors as missing energy. 
The spin difference may be used to discriminate the MUED model from other models 
\cite{Barr:2004ze}. However, the spin determination, particularly when the masses of 
new particles are degenerate, is difficult at the LHC, and hence the discrimination of 
TeV scale models is also difficult.

In order to discriminate the signals of the MUED model from those of other models, 
we focus on the existence of the KK tower. 
First of all, we need to speculate the value of $1/R$.
At the LHC, we can find the signals of first KK particles through 
the four lepton channel \cite{Brooijmans:2008se}. Through 
the observations of cascade decays of the first KK particles, we can 
speculate the value of $1/R$.
It is then possible to 
predict the masses of the second KK particles. 
The discovery of new particles with predicted masses strongly 
suggests that the MUED model is realized as new physics at the TeV scale.
At tree level, the second KK particles decay into lighter KK particles through KK 
number conserving processes, and eventually LKPs are left.
Since the LKP gives no signal at the detectors, it is very hard to measure the mass of
its parent particle. 
Fortunately, the second KK gauge bosons
directly couple with SM fermion pairs through KK number
violating interactions. Consequently, $\gamma^{(2)}$ and $Z^{(2)}$ 
decay into two charged leptons with nonzero branching ratios, 
and the masses of $\gamma^{(2)}$
and $Z^{(2)}$ can be clearly reconstructed from the dileptons
\cite{Datta:2005zs}. 
The mass difference between $\gamma^{(2)}$ and $Z^{(2)}$ is about 50 GeV for 
$1/R =$ 500 GeV, as shown in Table 1, and for this mass difference each resonance 
can be distinguished clearly by the observation of dielectron signals \cite{Datta:2005zs}. 
The double peak resonance also suggests the existence of the MUED model, 
because this is one of the typical signatures of this model.
For the discussion of the feasibility to confirm
the MUED model at the LHC, we need to know the event rate of the
dilepton signals. We thus calculate the production rates of $\gamma^{(2)}$
and $Z^{(2)}$ at the LHC in the following sections.

%%%%%%%%%%%%%%%%%%%%%%%%%%%%%%%%%%%%%%%%%%%%%%%%%%%%%%%%%%%
\section{Productions of $\gamma^{(2)}$ and $Z^{(2)}$}   %%%%%%%%%%%%%%%%%%%%%%%%%%%%%%
%%%%%%%%%%%%%%%%%%%%%%%%%%%%%%%%%%%%%%%%%%%%%%%%%%%%%%%%%%%

In this section, we discuss the Lagrangian relevant to the productions
of $\gamma^{(2)}$ and $Z^{(2)}$ bosons. With the Lagrangian we calculate 
the branching ratios of $g^{(2)}$ and $q^{(2)}$, which are necessary for the 
discussion of the indirect production of $\gamma^{(2)}$ and $Z^{(2)}$.

%%%%%%%%%%%%%%%%%%%%%%%%%%%%%%%%%%%%%%%%%%%%%%%%%%%%%%%%%%%%
\subsection{Lagrangian for $\gamma^{(2)}$ and $Z^{(2)}$ productions}      %%%%%%%%%%%%%%%%%%%%%%%
%%%%%%%%%%%%%%%%%%%%%%%%%%%%%%%%%%%%%%%%%%%%%%%%%%%%%%%%%%%%

First, we show the Lagrangian conserving the KK number relevant to gauge bosons, 
\begin{equation}
 \begin{split}
    \mathcal{L}_{\text{con}} 
    =
    &- g_i \sum_{n=1}^{\infty}
    \biggl[ 
          \bar f^{(n)} t^a \gamma^\mu f^{(n)} V_{i \mu}^{(0)a}    \\
    &~~~ + \bar f^{(n)} t^a \gamma^\mu P_{L(R)} f^{(0)} V_{i \mu}^{(n)a}
       + \bar f^{(0)} t^a \gamma^\mu P_{L(R)} f^{(n)} V_{i \mu}^{(n)a}   
    \biggr]   \\ 
    & -  \frac{g_i}{\sqrt{2}} \sum_{n, m=1}^{\infty}
    \biggl[ 
          \bar f^{(n)} t^a \gamma^\mu \gamma^5 f^{(m)} V_{i \mu}^{(n+m)a}    \\
    &~~~ + \bar f^{(n+m)} t^a \gamma^\mu f^{(n)} V_{i \mu}^{(m)a}
       + \bar f^{(n)} t^a \gamma^\mu f^{(n+m)} V_{i \mu}^{(m)a}    
    \biggr]   \\
    & + g_i f_i^{abc} \sum_{n=1}^{\infty}
    \biggl[
          (\partial_\mu V_{i \nu}^{(0)a}) V_i^{(n) b \mu} V_i^{(n) c \nu}   \\
       &~~~    
       + (\partial_\mu V_{i \nu}^{(n)a}) V_i^{(n) b \mu} V_i^{(0) c \nu}
       + (\partial_\mu V_{i \nu}^{(n)a}) V_i^{(0) b \mu} V_i^{(n) c \nu}   
    \biggr] ~,        \\
    %+  \frac{g_s}{\sqrt{2}}  f^{abc}
    %&\biggl[
    %      (\partial_\mu g_\nu^{(n+m)a}) g^{(n) b \mu} g^{(m) c \nu}   \\
    %   &    
    %   + (\partial_\mu g_\nu^{(n)a}) g^{(n+m) b \mu} g^{(m) c \nu}
    %   + (\partial_\mu g_\nu^{(n)a}) g^{(m) b \mu} g^{(n+m) c \nu}   
    %\biggr]    \\
    %+  g \varepsilon^{abc} \sum_{n=1}^{\infty}
    %&\biggl[
    %      (\partial_\mu W_\nu^{(0)a}) W^{(n) b \mu} W^{(n) c \nu}   \\
    %   &  
    %   + (\partial_\mu W_\nu^{(n)a}) W^{(n) b \mu} W^{(0) c \nu}
    %   + (\partial_\mu W_\nu^{(n)a}) W^{(0) b \mu} W^{(n) c \nu}   
    %\biggr]         ~,
    %+  \frac{g}{\sqrt{2}}  \varepsilon^{abc}
    %&\biggl[
    %      (\partial_\mu W_\nu^{(n+m)a}) W^{(n) b \mu} W^{(m) c \nu}  \\
    %   &
    %   + (\partial_\mu W_\nu^{(n)a}) W^{(n+m) b \mu} W^{(m) c \nu}
    %   + (\partial_\mu W_\nu^{(n)a}) W^{(m) b \mu} W^{(n+m) c \nu}   
    %\biggr] ~,
 \end{split}
\end{equation}
where the summation over $i$, $a$, $b$, and $c$ is implicitly made. 
$f^{(n)}$, $g_i$, $t^a$, and $V_{i \mu}^{a}$ are listed in Table 3. In the 
third part of the Lagrangian, $f_i^{abc}$ is the structure constant of 
$SU(3)$ for gluon and that of $SU(2)$ for the W boson. As long as we use 
the KK number conserving Lagrangian, second KK particles are produced 
in pair due to the KK number conservation, and hence the production 
rates are suppressed due to their small phase spaces.

\begin{table} [t!]
\begin{center}
\begin{tabular}{|l||c|c|c|}
\hline
  ~~~~~ n-th KK fermion $f^{(n)}$              & ~~~~~~$g_i$~~~~~~ & ~~~~~~$t^a$~~~~~~ & ~~~~~~$V_{i \mu}^{a}$~~~~~~ \\
\hline\hline
$SU(2)$-singlet charged lepton $E^{(n)}$   &   $- g'$              & $1$                    & $B_\mu$  \\
\hline
$SU(2)$-doublet lepton $L^{(n)}$              & $- (1/2) g'$       & $1$                    & $B_\mu$    \\
                                                         &        $g_2$        & $\sigma^a/2$       & $W^a_\mu$   \\
\hline
$SU(2)$-singlet up-type quark $U^{(n)}$    & $(2/3) g'$          & $1$                    & $B_\mu$ \\
                                                         &  $g_s$               & $\lambda^a/2$     & $g^a_\mu$      \\
\hline
$SU(2)$-singlet down-type quark $D^{(n)}$ & $- (1/3) g'$        & $1$                    & $B_\mu$ \\
                                                         &  $g_s$               & $\lambda^a/2$     &  $g^a_\mu$  \\
\hline
$SU(2)$-doublet quark $Q^{(n)}$              & $(1/6) g'$          & $1$                    & $B_\mu$ \\
                                                         &     $g_2$            &  $\sigma^a/2$     & $W^a_\mu$ \\
                                                         &     $g_s$            & $\lambda^a/2$     & $g^a_\mu$  \\
\hline
\end{tabular}
\caption{{\small{$f^{(n)}$, $g_i$, $t^a$, and $V_{i \mu}^{a}$ in the KK number conserving Lagrangian. 
                      $B$, $W$, and $g$ are $U(1)$, $SU(2)$, and $SU(3)$ gauge bosons, $g'$, $g_2$, and $g_s$ 
                      are $U(1)$, $SU(2)$, and $SU(3)$ gauge coupling constants, and $\sigma^a$ and $\lambda^a$ 
                      are Pauli matrices and Gell-Mann matrices, respectively.
            }} }
\label{table}
\end{center}
\end{table}

\begin{figure} [t!]
\begin{center}
\includegraphics[width=300pt,clip]{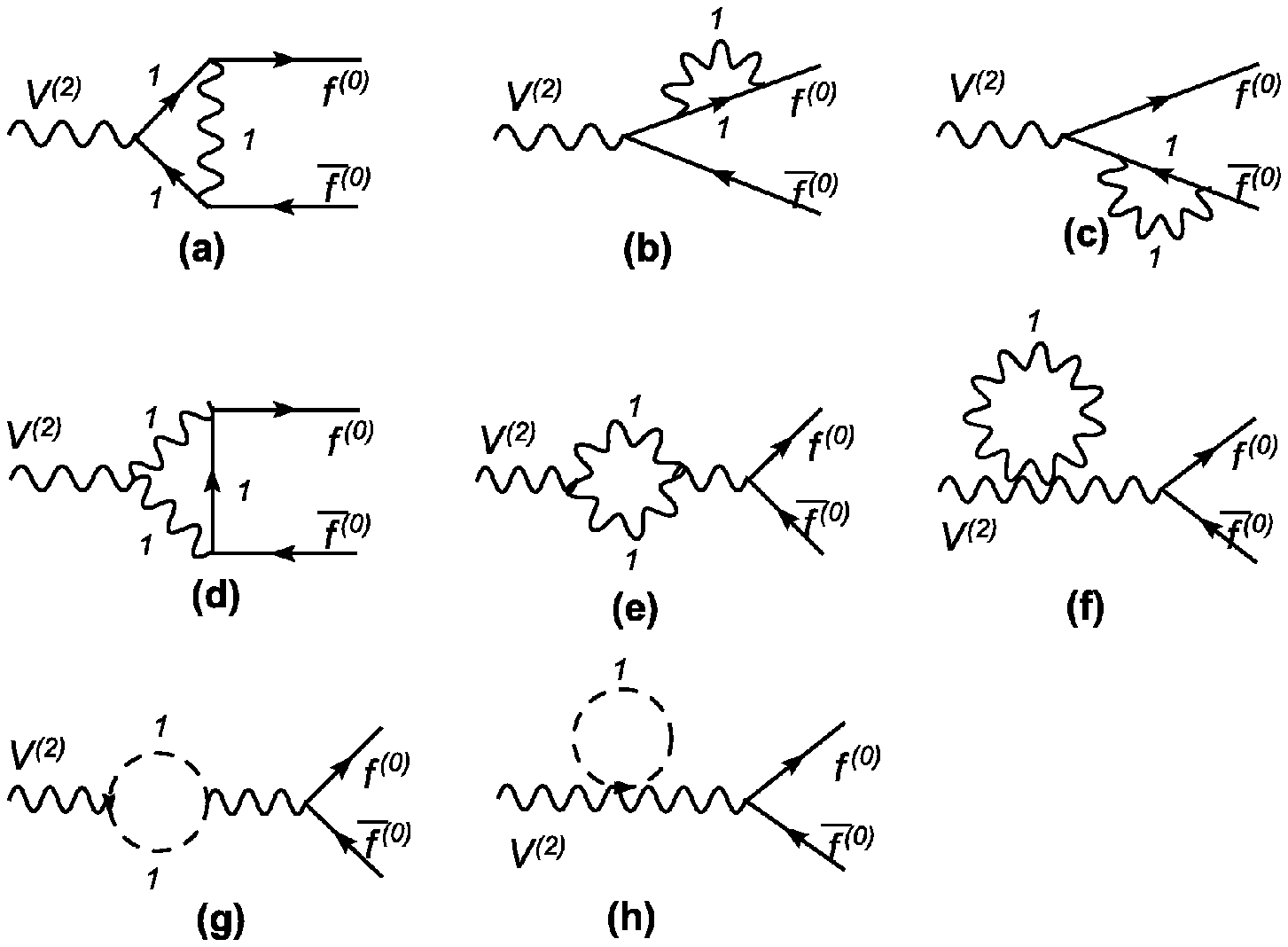}
\caption{{\small{The KK number violating vertices for $\bar f^{(0)} t^a \gamma^\mu P_{L (R)} f^{(0)} V_{i \mu}^{(2)a}$ 
            induced from gauge interactions. The attached number represents the KK number of a KK particle in the 
            loop.}} }
\label{loop}
\end{center}
%\end{figure} 
%\begin{figure} [h!]
\begin{center}
\includegraphics[width=290pt,clip]{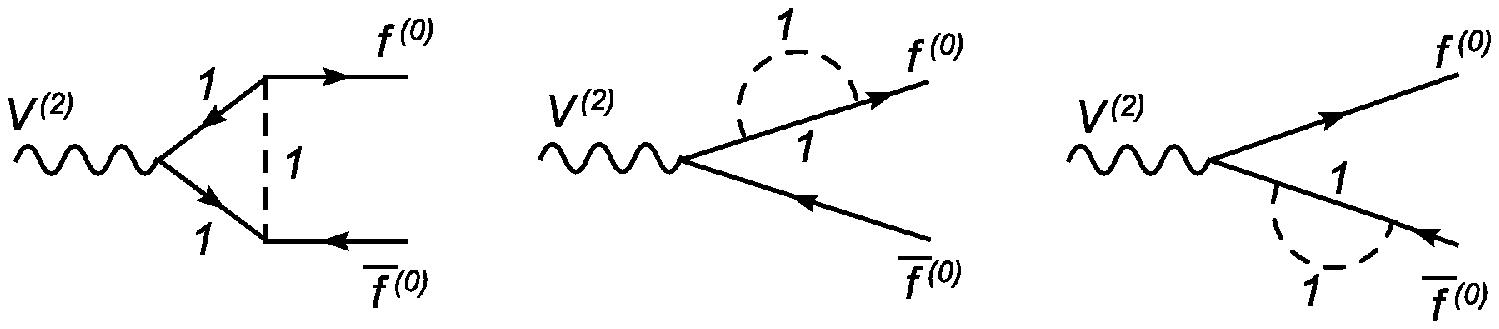}
\caption{{\small{The KK number violating vertices for $\bar f^{(0)} t^a \gamma^\mu P_{L (R)} f^{(0)} V_{i \mu}^{(2)a}$
            induced from Yukawa interactions.
            The attached number represents the KK number of a KK particle in the loop.}} }
\label{loop2}
\end{center}
\end{figure}

Next we discuss the KK number violating interactions. 
The KK number violating operators have been discussed in Ref. \cite{Cheng:2002iz} 
taking into account only the gauge interactions. In addition, we include also contributions of 
Yukawa interactions.
The effective Lagrangian for the KK number violating operators
turns out to be 
\begin{equation}
    \mathcal{L}_{\text{vio}}
    = 
    \frac{x_i}{4} 
    \Biggl\{
       ~N_i(f) c_t + 
       \left[ 
          9 C_j(f) - \frac{23}{3} C_j(G) \delta_{ij} + \frac{n_j}{3} \delta_{ij} 
       \right]
       c_j~
    \Biggr\}
    \bar f^{(0)} t_i^a \gamma^\mu P_{L (R)} f^{(0)} V_{i \mu}^{(2)a} ~,     \label{Lvio}
\end{equation}
\begin{equation}
 \begin{split}
    c_j \equiv \frac{ \sqrt{ 2 } x_j^2 }{ 16 \pi^2 } \text{log}\frac{ \Lambda^2 }{ \mu^2 } ~, ~~
    c_t \equiv \frac{ \sqrt{ 2 } y_t^2 }{ 16 \pi^2 } \text{log}\frac{ \Lambda^2 }{ \mu^2 }  ~.
 \end{split}       \label{c_i}
\end{equation}
Here, $y_t$ is the top Yukawa coupling constant, $x_i$, $n_i$, $C_i(f)$, $C_i(G)$, 
and $t_i^a$ are listed in Table 4, and $N_i(f)$ is listed in Table 5. Indices $i$, 
and $j$ run over the SM gauge interactions $U(1)$, $SU(2)$, and $SU(3)$, and 
summation over $f$ is implicitly made. The renormalization scale is denoted by 
$\mu$. Contribution $9 C_j(f)$ comes from Figs. \ref{loop}(a) - \ref{loop}(c), 
contribution $- (23/3) C_j(G) \delta_{ij}$ comes from Figs. \ref{loop}(d) -  
\ref{loop}(f), and contribution $(n_j/3) \delta_{ij}$ comes from Figs. \ref{loop}(g) 
and \ref{loop}(h). Contribution $N_i(f) c_t$ comes from diagrams in Fig. \ref{loop2}.

\begin{table} [!ht]
\begin{center}
\begin{tabular}{|c||c|c|c|}
\hline
~~~~~~~~~~~& ~~~$U(1)$~~~ & ~~~$SU(2)$~~~ & ~~~$SU(3)$~~~ \\
\hline\hline
$x_i$ & $g' Y_f$ & $g_2$ & $g_s$  \\
\hline
$n_i$ & 1 & 2 & 0 \\
\hline
$C_j(f)$ & $Y_f^2$ & 3/4 & 4/3 \\
\hline
$C_i(G)$ & 0 & 2 & 3 \\
\hline
$t_i^a$ & \textbf{1} & $\sigma^a/2$ & $\lambda^a/2$ \\
\hline
\end{tabular}
\caption{{\small{Coefficient in the KK number violating operator in the effective Lagrangian [Eq. (\ref{Lvio})]. $g'$, $g_2$, 
                      and $g_s$ are $U(1)$, $SU(2)$, and $SU(3)$ gauge coupling constants, and $Y_f$ is 
                      $U(1)$ hypercharge. $\sigma^a$ and $\lambda^a$ are Pauli and Gell-Mann matrices, respectively.
           }} }
\label{tab:a}
\end{center}
\end{table}

\begin{table} [!ht]
\begin{center}
\begin{tabular}{|c||c|c|c|c|c|c|}
\hline
~~~& $Q_3^{(0)} Q_3^{(0)} \gamma^{(2)}$, $Q_3^{(0)} Q_3^{(0)} W^{(2)}$, $Q_3^{(0)} Q_3^{(0)} g^{(2)}$, $T^{(0)} T^{(0)} g^{(2)}$
     & $T^{(0)} T^{(0)} \gamma^{(2)}$
     & Other         \\
\hline\hline
$N(f)$     & ~~$1$~~ & ~~$5$~~ & 0   \\
\hline
\end{tabular}
\caption{{\small{Coefficient $N(f)$ in the KK number violating operator in the effective Lagrangian [Eq. (\ref{Lvio})]. $Q_3$ 
                      is the third generation $SU(2)$-doublet quark, and $T$ is the $SU(2)$ singlet top quark.
           }} }
\label{tab:b}
\end{center}
\end{table}

%%%%%%%%%%%%%%%%%%%%%%%%%%%%%%%%%%%%%%%%%%%%%%%%%%%%%%%%%%%%%%%%%%%%%%%%%%%%%%%%%%%%%%%%%%%%%%%%
\subsection{Production processes of $\gamma^{(2)}$ and $Z^{(2)}$} %%%%%%%%%%%%%%%%%%%%%%%%%%%%%%%%%%%%%%%%%%
%%%%%%%%%%%%%%%%%%%%%%%%%%%%%%%%%%%%%%%%%%%%%%%%%%%%%%%%%%%%%%%%%%%%%%%%%%%%%%%%%%%%%%%%%%%%%%%%

We are now in a position to discuss the production processes of
$\gamma^{(2)}$ and $Z^{(2)}$. At the LHC, $\gamma^{(2)}$ and $Z^{(2)}$
are produced through two type of processes : (1) direct productions and (2)
indirect productions via the cascade decays of the second KK colored
particles. The production cross sections of $\gamma^{(2)}$ and $Z^{(2)}$ have 
originally been calculated in Ref. \cite{Datta:2005zs}, and their calculation 
includes all of the KK number conserving processes and the direct one-body 
production processes of the second KK gauge bosons. In this article, we 
calculate the production cross sections of $\gamma^{(2)}$ and $Z^{(2)}$ 
including all significant processes. For example, our calculation includes 
$pp \rightarrow q^{(2)} q^{(0)}$, $pp \rightarrow \gamma^{(2)} q^{(0)}$, 
$pp \rightarrow q^{(2)} \bar q^{(0)}$, and so on. Importantly, these 
processes provide large contributions to $\gamma^{(2)}$ and $Z^{(2)}$ 
productions, particularly for large $1/R$ ($\gtrsim$ 800 GeV). We show the 
relevant processes to the $\gamma^{(2)}$ production in Figs. 3 - 8.

\begin{figure} [p!]
\begin{center}
\includegraphics[width=300pt, clip]{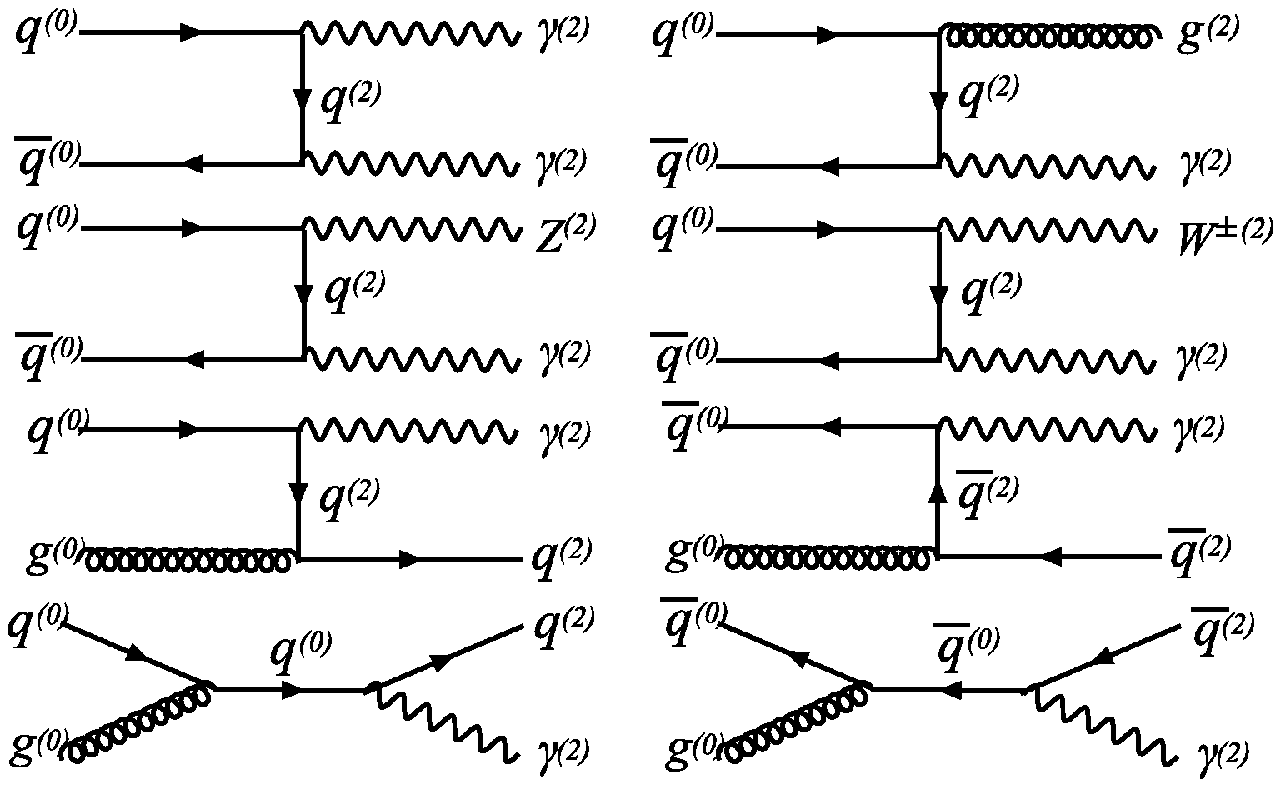}
\caption{{\small{The production of $\gamma^{(2)}$ through KK number conserving processes.}} }
\label{gamma2_pro_conserving}
\end{center}
%\end{figure}
%\begin{figure} [h!]
\begin{center}
\includegraphics[width=430pt,clip]{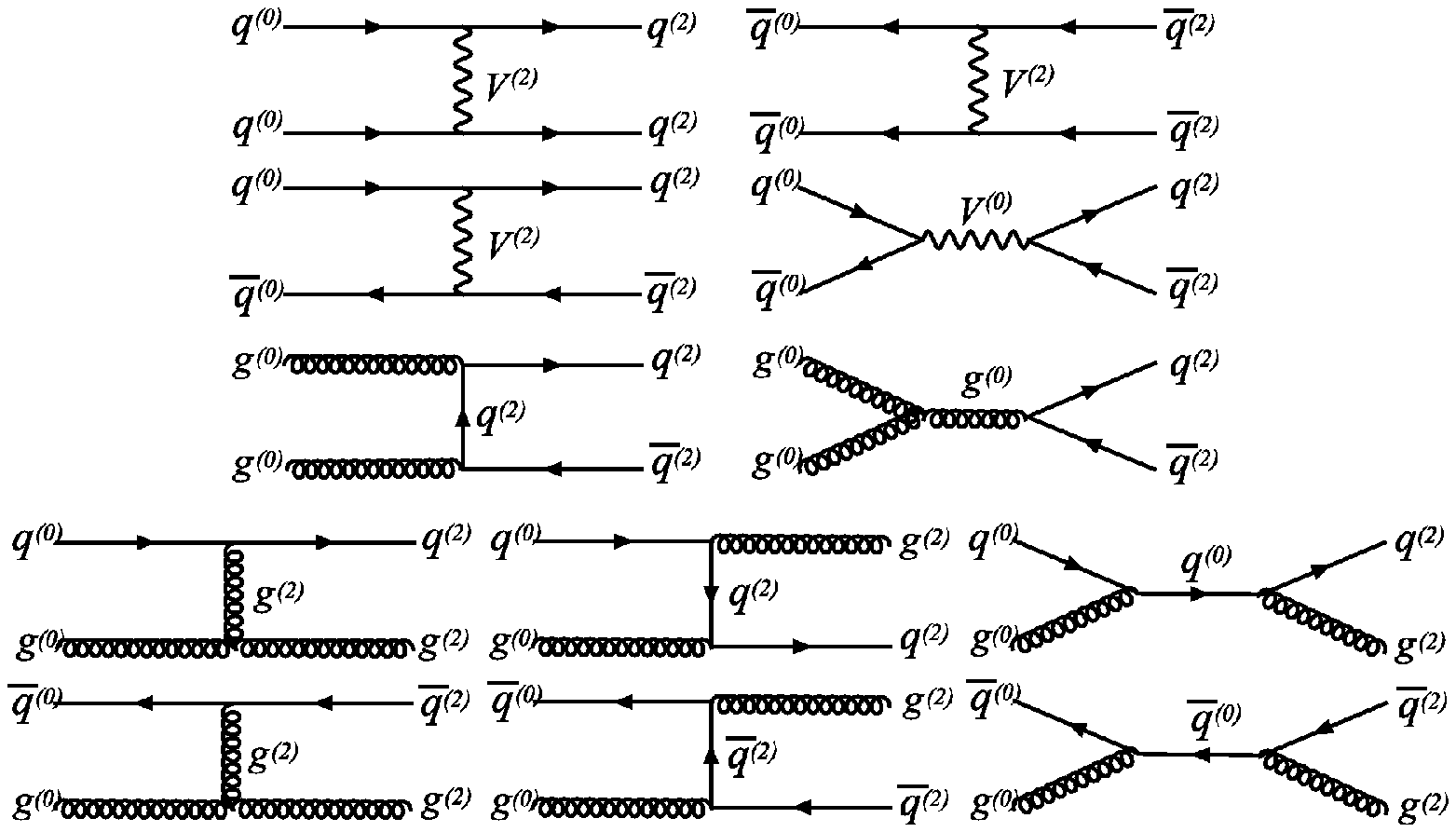}
\caption{{\small{The production of $q^{(2)}$ through KK number conserving processes.
            $V$ stands for $\gamma$, $W^{\pm }$, $Z$, and $g$. }} }
\label{q2_pro_conserving}
\end{center}
%\end{figure}
%\begin{figure} [t!]
\begin{center}
\includegraphics[width=430pt,clip]{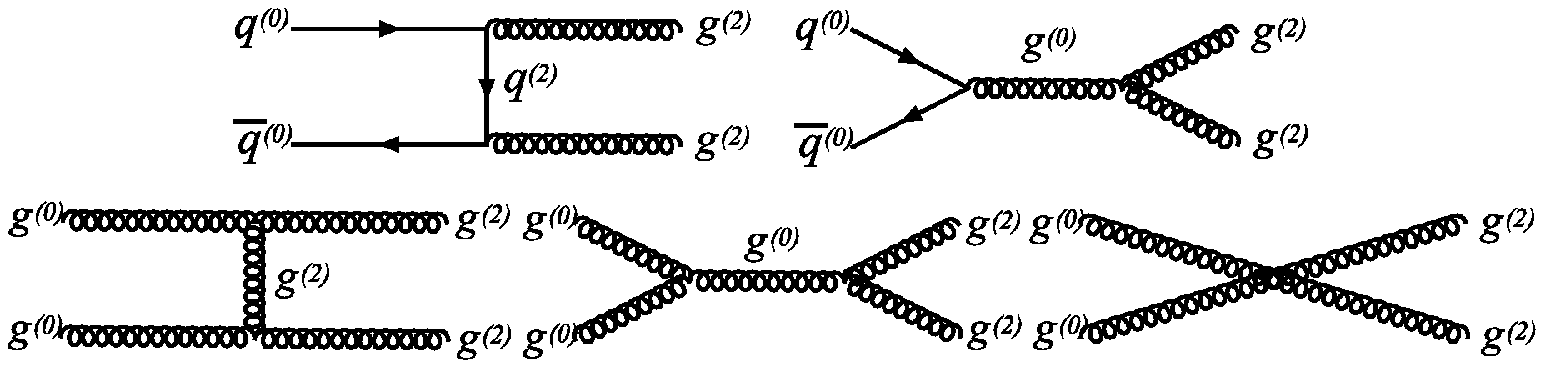}
\caption{{\small{The production of $g^{(2)}$ through KK number conserving processes. }} }
\label{g2_pro_conserving}
\end{center}
\end{figure}

\begin{figure} [t!]
\begin{center}
\includegraphics[width=305pt,clip]{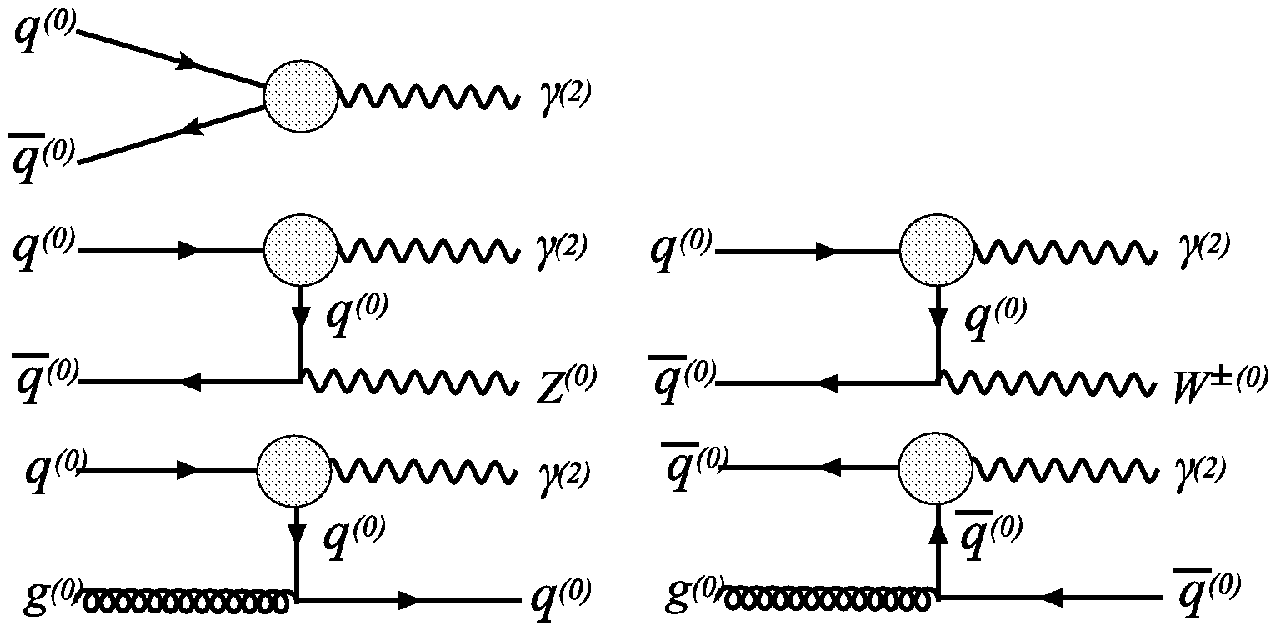}
\caption{{\small{The direct production of  $\gamma^{(2)}$ through KK number violating processes.
                      The gray circle represents the KK number violating vertex.
                      }} }
\label{qq_to_gamma2}
\end{center}
%\end{figure}
%\begin{figure} [h!]
\begin{center}
\includegraphics[width=310pt,clip]{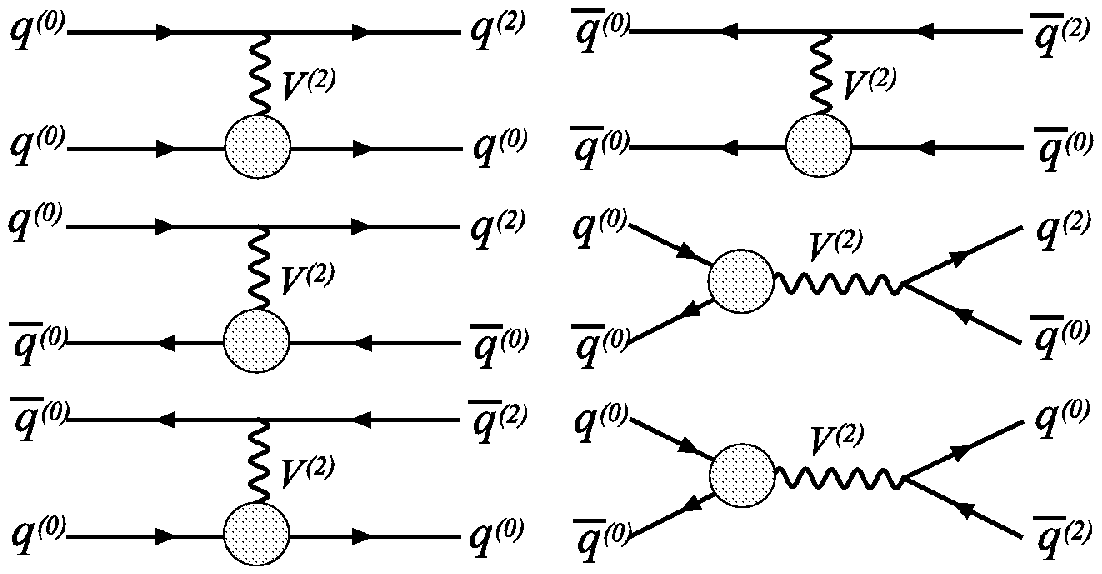}
\caption{{\small{The production of $q^{(2)}$ through KK number violating processes.
            $V^{(2)}$ stands for $\gamma^{(2)}$, $W^{\pm (2)}$, $Z^{(2)}$, and $g^{(2)}$.
            The gray circle represents the KK number violating vertex.
            }} }
\label{q2_pro_violating}
\end{center}
%\end{figure}
%\begin{figure} [h!]
\begin{center}
\includegraphics[width=425pt,clip]{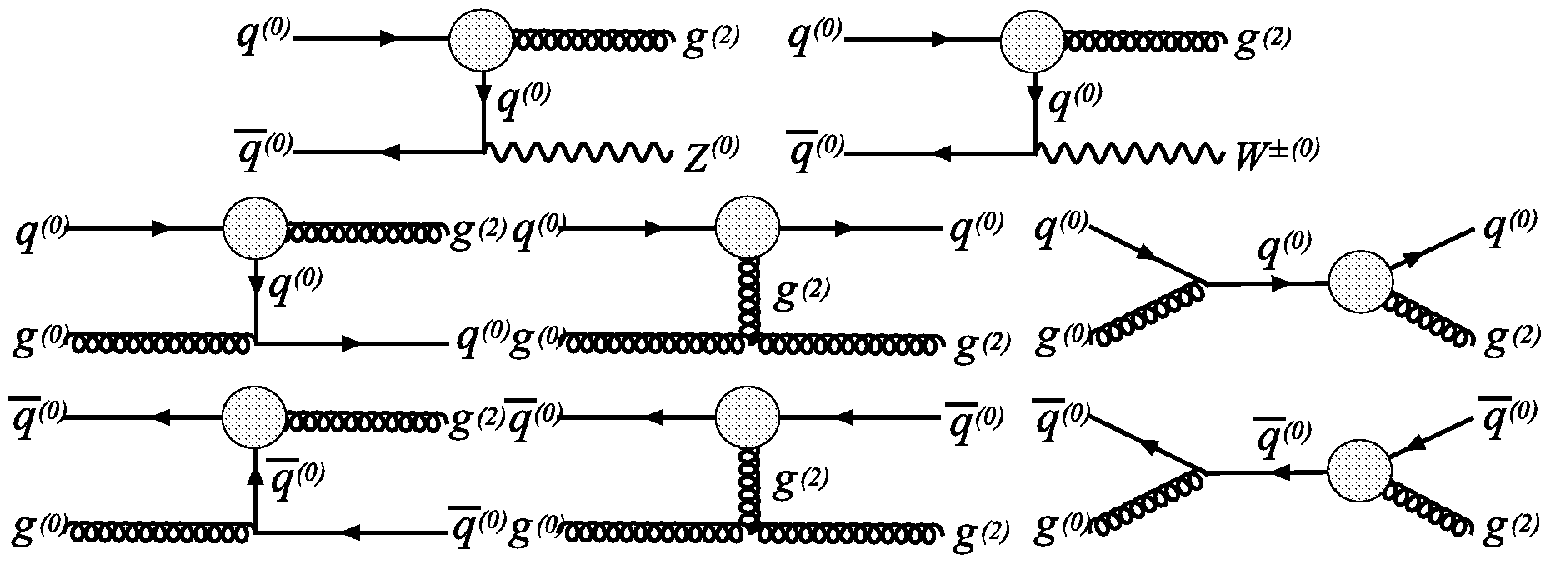}
\caption{{\small{The production of $g^{(2)}$ through KK number violating processes.
            The gray circle represents the KK number violating vertex.
            }} }
\label{g2_pro_violating}
\end{center}
\end{figure}

Figure 3 (Fig. 6) shows the direct production of $\gamma^{(2)}$ through
KK number conserving (violating) processes. Some of these processes 
also produce $q^{(2)}$, $\bar q^{(2)}$, and $g^{(2)}$, and 
they decay into $\gamma^{(2)}$ and $Z^{(2)}$. We must 
include those contributions in the calculation of
the production cross section of $\gamma^{(2)}$ and $Z^{(2)}$. 
Figure 4 (Fig. 7) shows the production of $q^{(2)}$ and $\bar q^{(2)}$
through KK number conserving (violating) processes, and Fig. 5
(Fig. 8) shows the production of $g^{(2)}$ through KK number
conserving (violating) processes. These colored particles decay 
into $\gamma^{(2)}$ and $Z^{(2)}$. Here the gray circle  
in Figs. 6 - 8 stands for the KK number violating vertex.
In the $s$-channel processes in Fig. 7, the contributions from $g^{(2)}$ 
one-body direct production are included, while the contributions from 
$\gamma^{(2)}$ one-body direct production are not included.
Hence, $\gamma^{(2)}$ one-body direct 
production is shown in Fig. 6, and $g^{(2)}$ one-body direct production
is not shown in Fig. 8. Note that the $\gamma^{(2)}$ 
production from the $Z^{(2)}$ decay can be neglected, because
the branching ratio $Z^{(2)} \rightarrow L^{(2)} \bar L^{(0)}
\rightarrow \gamma^{(2)} L^{(0)} \bar L^{(0)}$ is 
small enough. The processes of $Z^{(2)}$ production are almost the same as
the $\gamma^{(2)}$ production, so we can skip the discussion of the
$Z^{(2)}$ production.
In Figs. 6 and 8, processes which have the final state 
$\gamma^{(2)} \gamma^{(0)}$, $\gamma^{(2)} g^{(0)}$, 
$g^{(2)} \gamma^{(0)}$, or $g^{(2)} g^{(0)}$ are not shown, because 
in this calculation, $\gamma^{(0)}$ and $g^{(0)}$ in these processes 
are the origin of the infrared divergences for the one-body production processes 
of second KK gauge bosons. The divergences can be removed by the calculation 
with a complete treatment.

\begin{figure}[t!]
 \begin{minipage}{0.5\hsize}
  \begin{center}
   \includegraphics[width=205pt, angle=0, clip]{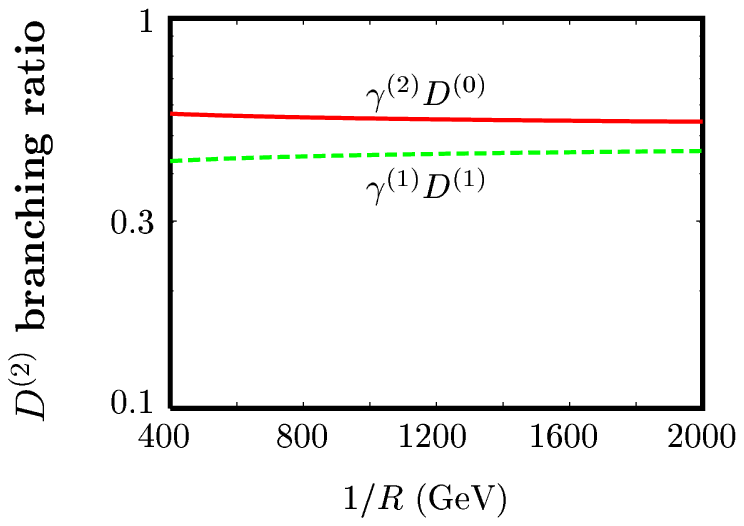}
  \end{center}
  \caption{{\small{The branching ratio of $D^{(2)}$.}} }
  \label{D2BR}
 \end{minipage}   \hspace{2mm}
 \begin{minipage}{0.5\hsize}
  \begin{center}
   \includegraphics[width=205pt, angle=0, clip]{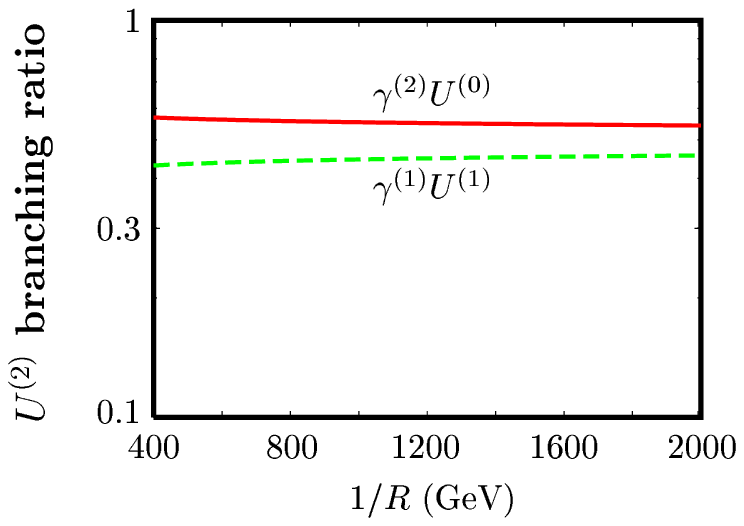}
  \end{center}
  \caption{{\small{The branching ratio of $U^{(2)}$.}} }
  \label{U2BR}
 \end{minipage}
%\end{figure}  
%\begin{figure}[t!]
 \begin{minipage}{0.50\hsize}
  \begin{center}
   \includegraphics[width=195pt, angle=0, clip]{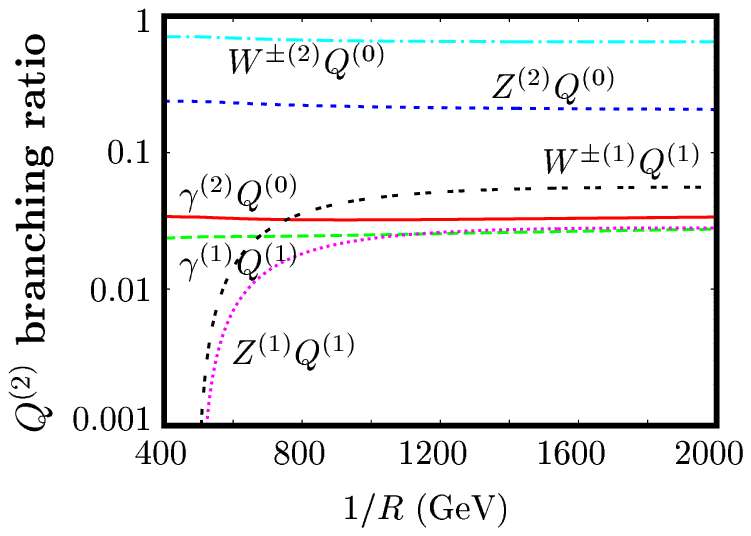}
  \end{center}
  \caption{{\small{The branching ratio of $Q^{(2)}$.}} }
  \label{Q2BR}
 \end{minipage}   \hspace{2mm}
 \begin{minipage}{0.5\hsize}
 \begin{center}
  \includegraphics[width=205pt, angle=0]{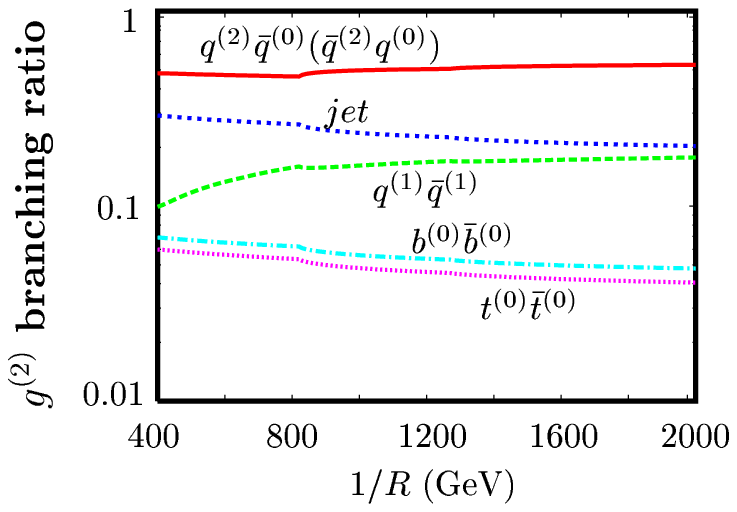}
  \caption{{\small{The branching ratio of $g^{(2)}$.}}}
 \label{g2BR}           
 \end{center}
 \end{minipage}
\end{figure}

In order to calculate the indirect production cross sections of $\gamma^{(2)}$ and 
$Z^{(2)}$, we calculate the branching ratio of $g^{(2)}$ and $q^{(2)}$. We show the 
branching ratio of the $SU(2)$ singlet KK down-type quark $D^{(2)}$ in Fig. \ref{D2BR}, 
the branching ratio of the $SU(2)$ singlet KK up-type quark $U^{(2)}$ in 
Fig. \ref{U2BR}, and the branching ratio of the $SU(2)$ doublet KK quark $Q^{(2)}$ in 
Fig. \ref{Q2BR} as a function of $1/R$. Since $U^{(2)}$ and $D^{(2)}$ are $SU(2)$ 
singlet, they couple only with the $U(1)$ hypercharge gauge boson. 
Please note that 
the Weinberg angle of the neutral KK gauge boson is almost equal to 0 
\cite{Cheng:2002iz}, and hence it is possible to identify the KK hypercharge gauge boson 
as the KK photon. 
Thus, as shown in Figs. \ref{D2BR} and \ref{U2BR}, $D^{(2)}$ and 
$U^{(2)}$ decay into the KK photon, and $D^{(2)}$ and $U^{(2)}$ can be dominant 
sources of $\gamma^{(2)}$. On the other hand, $Q^{(2)}$ dominantly decays into 
$Z^{(2)}$, and hence $Q^{(2)}$ are dominant sources of $Z^{(2)}$.

Finally, we show the branching ratio of $g^{(2)}$ in Fig. \ref{g2BR}. In this figure, 
generation indices are implicitly summed for each line, while $jet$ represents the sum of 
first and second generation SM quarks. At the LHC, the production rates of the 
second KK colored particles are much larger than those of other second KK particles, 
and $g^{(2)}$ and $q^{(2)}$ dominantly decay into a second KK particle and a SM 
particle, as shown in these figures. Hence, the indirect productions of $\gamma^{(2)}$ 
and $Z^{(2)}$ from their cascade decays are quite significant. 
%%%
Note that the branching ratios of each second KK particle calculated in this work are 
different from that in Ref. \cite{Datta:2005zs} due to two reasons. The one is the 
difference of the KK number violating operators. In particular, we include the KK 
number violating operators induced by the top Yukawa coupling. 
%%%
The other is the difference of mass spectrum of KK particles. By comparing 
our mass spectrum (Table 2) with theirs (Fig. 1 in Ref. \cite{Datta:2005zs}), 
we find the difference in these mass spectra.
%%%
The difference of the branching ratios from the previous work \cite{Datta:2005zs} 
mainly arises from the latter reason. 
%%%
In the next section, we will calculate the production rates of $\gamma^{(2)}$ and $Z^{(2)}$ 
using the results in this section.

%\clearpage
%%%%%%%%%%%%%%%%%%%%%%%%%%%%%%%%%%%%%%%%%%%%%%%%%%
\section{Numerical results} %of $\gamma^{(2)}$ and $Z^{(2)}$ production cross sections}%%%%%%
%%%%%%%%%%%%%%%%%%%%%%%%%%%%%%%%%%%%%%%%%%%%%%%%%%

In this section, we present numerical results of the cross sections
for $\gamma^{(2)}$ and $Z^{(2)}$ productions and estimate the number of
the dilepton signal from $\gamma^{(2)}$ and $Z^{(2)}$ decays at the
LHC. The calculations of the cross sections have been performed by using
the calcHEP \cite{Pukhov:2004ca} implementing the Lagrangian Eq. (\ref{Lvio})
derived in the previous section.

\begin{figure} [!ht]
\begin{center}
\includegraphics[width=385pt, angle=0,clip]{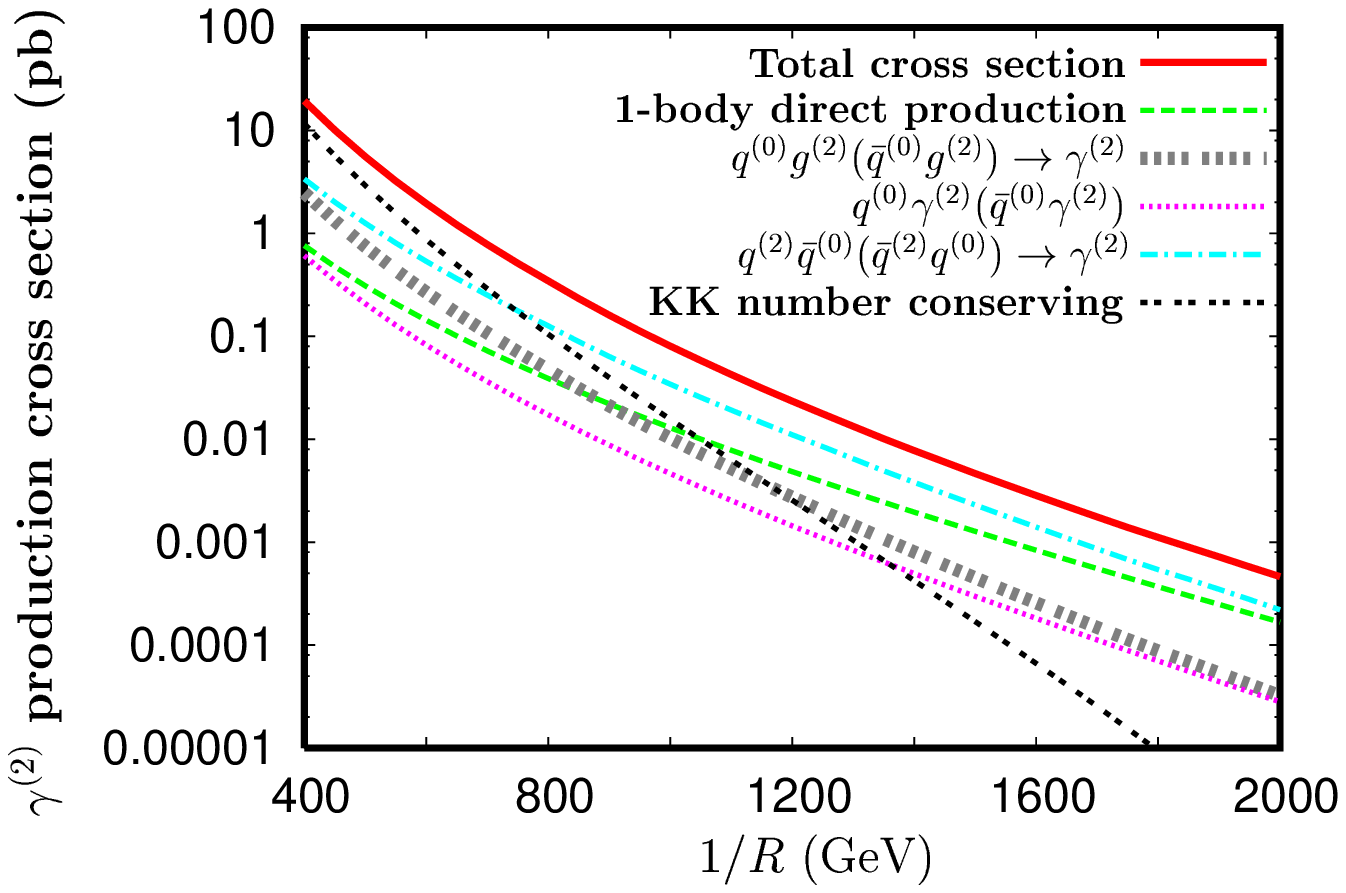}
\caption{{\small{The production cross sections of $\gamma^{(2)}$. 
The red solid line shows the total cross section, and 
other lines show the production cross sections of $\gamma^{(2)}$ 
for each process as denoted in the legend.}}}
\label{gamma2_each_process}
\end{center}
%\end{figure}
%\begin{figure} [!ht]
\begin{center}
\includegraphics[width=385pt, angle=0,clip]{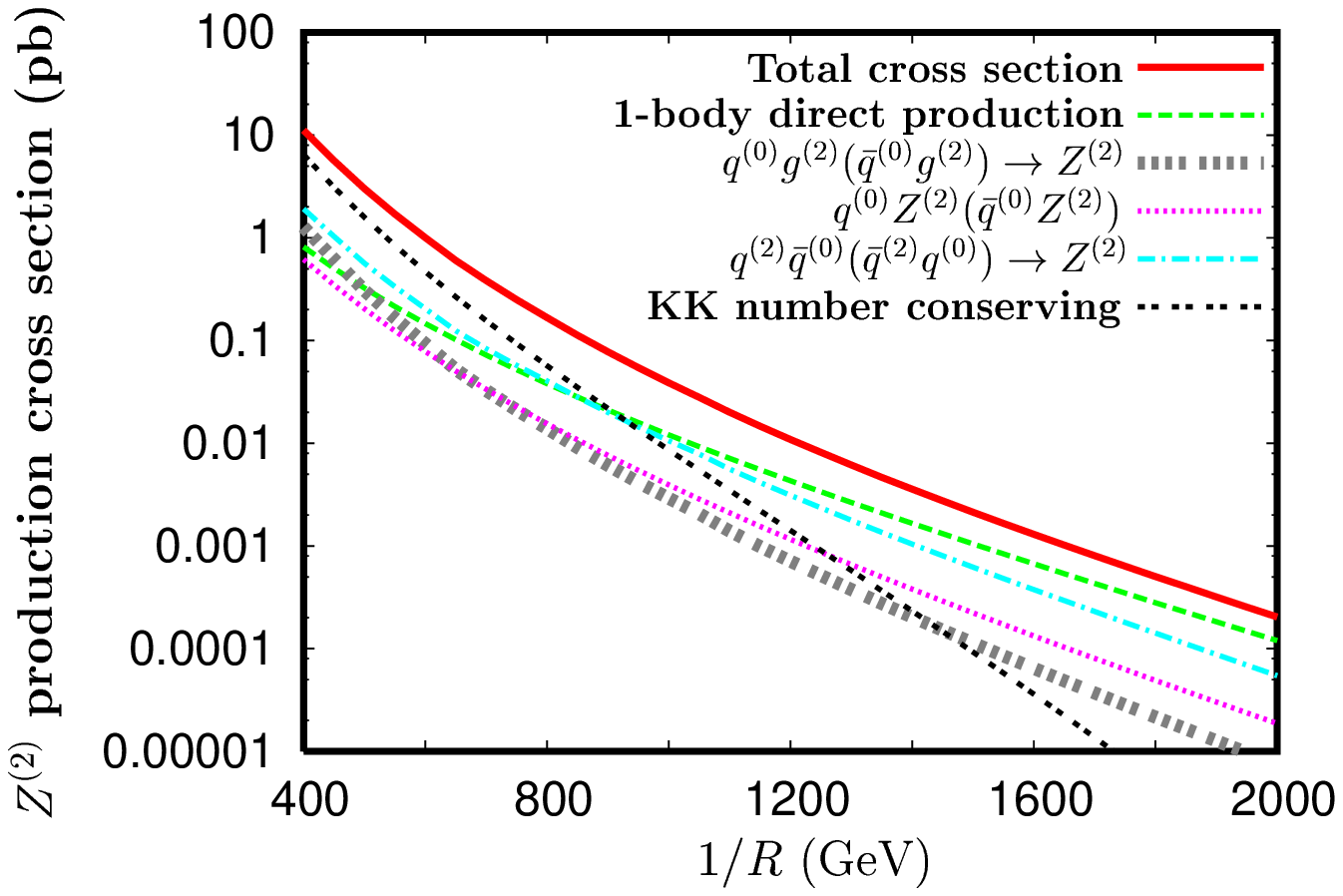}
\caption{{\small{The production cross sections of $Z^{(2)}$. 
The red solid line shows the total cross section, and 
other lines show the production cross sections of $Z^{(2)}$ 
for each process as denoted in the legend.}}}
\label{Z2_each_process}
\end{center}
\end{figure}
In Fig. \ref{gamma2_each_process} (Fig. \ref{Z2_each_process}), we show the
production cross section of $\gamma^{(2)}$ ($Z^{(2)}$) as a function of
$1/R$. In the calculation, we have used the CTEQ6L code
\cite{Pumplin:2002vw} as a parton distribution function (PDF). The red solid
line shows the total cross section of $\gamma^{(2)}$ ($Z^{(2)}$)
production. 
%%%
Note that although we show the results for only significant processes in these 
figures and discuss them, all processes shown in Figs. \ref{gamma2_pro_conserving} 
- \ref{g2_pro_violating} are included in the calculation of the total cross section.
%%%
As shown in these figures, for large $1/R$ ($\gtrsim$ 800
GeV), the KK number violating processes dominantly contribute to the total
production cross section. This means that the KK number violating operators
are important for the discrimination of the MUED model from other TeV
scale models.

We discuss the significance of each production process. 
(i) For large $1/R$ ($1/R \gtrsim$ 800 GeV), one-body direct production
processes are important, because even if $\gamma^{(2)}$ and $Z^{(2)}$ are
too heavy to be produced in pair, these processes provide $\gamma^{(2)}$ and 
$Z^{(2)}$ efficiently. 
(ii) In addition to evading the suppression of small phase space, the processes 
$pp \rightarrow q^{(0)} V^{(2)} (\bar q^{(0)} V^{(2)})$ have notable features.
Since the PDF of proton is dominated by gluons and the valence quarks, i.e., up- 
and down-quarks, the rate of $q^{(0)} g^{(0)}$ collision is larger than  
$q^{(0)} \bar q^{(0)}$ collision. Furthermore, the 
cross section of these processes has the large logarithm factor :
\begin{equation}
 \begin{split}
   \sigma \sim \text{log} \biggl[ \frac{ (s - m_{V2}^2) + m_{q0}^2 }{ m_{q0}^2 } \biggr] .
 \end{split}
\end{equation}
Here $m_{q0}$ is the mass of a SM quark, $m_{V2}$ is the mass of the second KK gauge 
boson, and $s$ is the square of the initial state total energy. In usual cases, cross 
sections decrease according to the increasing of $1/R$ and $s$. However, this logarithm 
factor prevents the drastic decreasing. 
%%%
Thus, in all ranges of $1/R$ in these figures, indirect productions of $\gamma^{(2)}$ 
and $Z^{(2)}$ from $pp \rightarrow q^{(0)} g^{(2)} (\bar q^{(0)} g^{(2)})$ and direct 
productions $pp \rightarrow q^{(0)} \gamma^{(2)} (\bar q^{(0)} \gamma^{(2)} )$ or 
$pp \rightarrow q^{(0)} Z^{(2)} (\bar q^{(0)} Z^{(2)} )$ provide a non-negligible contribution 
to their production cross sections.
%%%
(iii) At the LHC, the final state $q^{(2)} \bar q^{(0)}$ (or $\bar q^{(2)} q^{(0)}$) is 
mostly provided by the $s$-channel process mediated by the $g^{(2)}$ propagator. In other 
words, these include the contributions from $g^{(2)}$ one-body production and indirect 
productions through the cascade decay of $g^{(2)}$.
In this case, the pole resonance of $g^{(2)}$
leads the large enhancement of the cross section. Thus the processes $pp
\rightarrow q^{(2)} \bar q^{(0)} ~(\bar q^{(2)} q^{(0)})$
have large cross sections, even if $1/R$ is rather large, and provide 
large contributions to the indirect productions of $\gamma^{(2)}$ and $Z^{(2)}$. 
(iv) For small $1/R$ ($\lesssim$ 800 GeV), the cross sections of KK number 
conserving processes are larger than those of KK number violating processes, 
because the KK number conserving operators have no loop suppressions.  However, 
for large $1/R$ ($\gtrsim$ 800 GeV), the contribution from KK number conserving 
processes decreases rapidly because of the severe kinematical suppression on the 
pair production of the second KK particles.

\begin{figure}[t!]
 \begin{minipage}{0.49\hsize}
  \begin{center}
   \includegraphics[width=215pt, angle=0,clip]{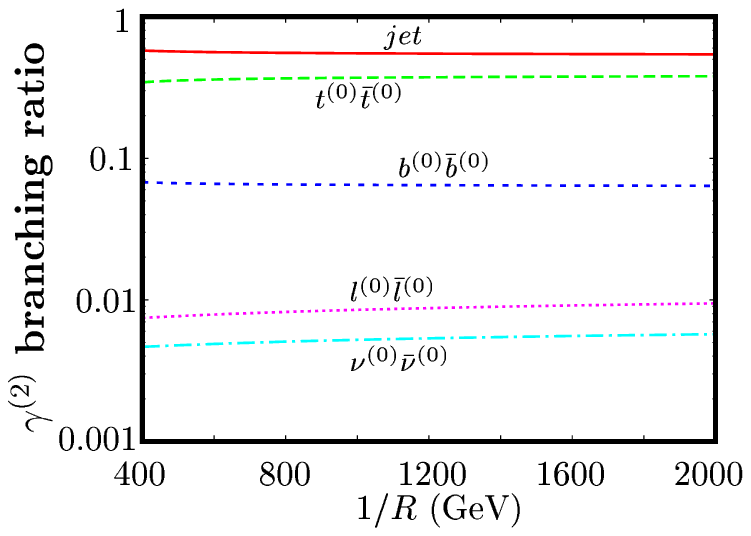}
  \end{center}
  \caption{{\small{The branching ratio of $\gamma^{(2)}$.}}}
  \label{gamma2BR}
 \end{minipage}  \hspace{2mm}
 \begin{minipage}{0.49\hsize}
  \begin{center}
   \includegraphics[width=215pt, angle=0]{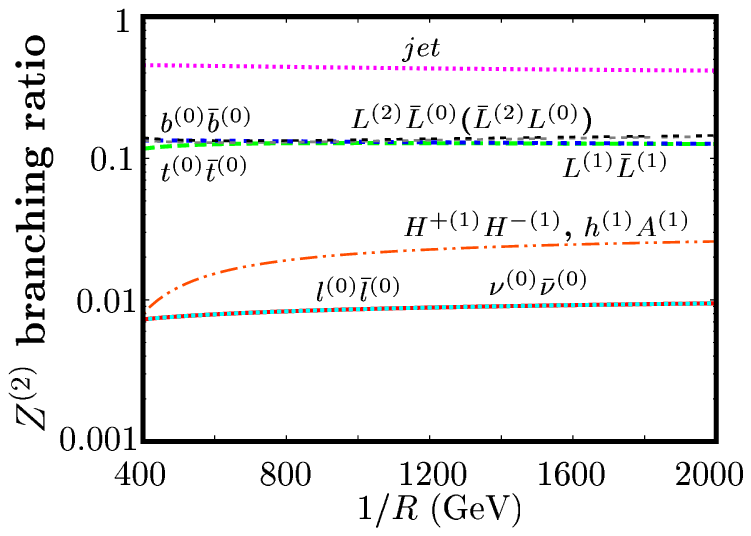}
  \end{center}
  \caption{{\small{The branching ratio of $Z^{(2)}$.}}}
  \label{Z2BR}
 \end{minipage}
\end{figure}

Assuming an integrated luminosity of 100 fb$^{-1}$, the number of produced 
$\gamma^{(2)}$ and $Z^{(2)}$ are calculated as 10$^6$ - 10$^2$ for 400 GeV 
$\leq 1/R \leq$ 2000 GeV. Once $\gamma^{(2)}$ and $Z^{(2)}$ are produced, 
they decay into dileptons with nonzero 
branching ratios. In order to estimate the number of dilepton signals from 
$\gamma^{(2)}$ and $Z^{(2)}$, we calculate these branching ratios. Figure 
\ref{gamma2BR} (Fig. \ref{Z2BR}) shows the branching ratio of $\gamma^{(2)}$ 
($Z^{(2)}$). In both figures, each line generation indices is implicitly summed.  
$\gamma^{(2)}$ is the lightest second KK particle, and the masses of the first KK 
fermions are always heavier than half of the $\gamma^{(2)}$ mass. Then 
$\gamma^{(2)}$ must decay into a SM fermion pair through KK number violating 
processes. On the other hand, $Z^{(2)}$ has many decay channels comparing with 
$\gamma^{(2)}$. Although the decay channels of $\gamma^{(2)}$ and $Z^{(2)}$ are 
quite different with each other, coincidentally the branching ratio to dilepton is almost 
the same.

In Table \ref{dilepton}, we show the number of the dilepton signals assuming the 
luminosity 100 fb$^{-1}$. 
%%%
Table \ref{dilepton} shows that for $1/R \lesssim 1600$ GeV it remains possible   
that we can discriminate the MUED model from other models by using the dilepton signals with 
100 fb$^{-1}$ integrated luminosity.
%%%
If the MUED model is realized by the nature, in addition to the dilepton signals from 
the decay of $\gamma^{(2)}$ and $Z^{(2)}$, many new particles with degenerate 
mass spectrum around $1/R$ (i.e., the first KK particles) are discovered. Connecting 
the observational results of dilepton signals and the discovery of the first KK particles, 
it is possible to confirm the MUED model. 
In this work, we do not discuss the feasibility of the MUED model confirmation, 
which needs a complete analysis with Monte Carlo simulation. It is beyond the scope of 
this paper. This will be addressed in future work \cite{2009}.

\begin{table} [t!]
\begin{center}
\begin{tabular}{|l||l|l|l|}
\hline
~~~$1/R$~~~& Dileptons from $\gamma^{(2)}$ & Dileptons from $Z^{(2)}$  \\
\hline\hline
400 GeV & 1.5 $\times $ 10$^4$ & 9.4 $\times $ 10$^3$   \\
\hline
800 GeV & 2.9 $\times $ 10$^2$ & 1.6 $\times $ 10$^2$  \\
\hline
1200 GeV & 2.1 $\times $ 10 & 1.2 $\times $ 10  \\
\hline
1600 GeV & 2.6 & 1.4  \\
\hline
2000 GeV & 4.4 $\times $ 10$^{-1}$ & 2.3 $\times $ 10$^{-1}$  \\
\hline
\end{tabular}
\caption{{\small{The number of the dilepton signals at the LHC with 100 fb$^{-1}$.
           }} }
\label{dilepton}
\end{center}
\end{table}

%%%%%%%%%%%%%%%%%%%%%%%%%%%%%%%%%%%%%%%%%%%%%%%%%%%%%%%%%%%%
\section{Summary} %%%%%%%%%%%%%%%%%%%%%%%%%%%%%%%%%%%%%%%%%%%%%%%%%%
%%%%%%%%%%%%%%%%%%%%%%%%%%%%%%%%%%%%%%%%%%%%%%%%%%%%%%%%%%%%

At the LHC, the discrimination of the MUED model from other models is difficult, 
because the signals of new particles of TeV scale models are quite similar to each 
other. For the discrimination, we focused on a distinct feature of the MUED model, 
i.e., the existence of the KK tower. Once $\gamma^{(2)}$ and $Z^{(2)}$ are 
produced, they can decay into dilepton which provides a very clear signal of the 
second KK particles. The discovery of the second KK particles strongly indicates 
the existence of the KK tower and consequently will lead to the confirmation of the 
MUED model. In order to estimate the number of the dilepton events from 
$\gamma^{(2)}$ and $Z^{(2)}$, we have calculated the production rates of 
$\gamma^{(2)}$ and $Z^{(2)}$ at the LHC.

First we have calculated the KK number violating operators. They play a crucial 
role for the discrimination, because $\gamma^{(2)}$ and $Z^{(2)}$ can decay 
into dilepton through these couplings.
%%%
In the calculation, we have included the contribution of the top Yukawa 
coupling to improve the KK number violating operators.
%%%
This have improved the branching ratio calculations of the second KK gauge bosons.
%%%
Next we have shown all significant processes for $\gamma^{(2)}$ and $Z^{(2)}$ 
productions including both the KK number conserving and violating interactions. 
Then we have calculated the production cross sections of $\gamma^{(2)}$ 
and $Z^{(2)}$. Finally we have found that $\cal {O}$(10$^6$) - $\cal {O}$(10$^2$) of 
$\gamma^{(2)}$ and $Z^{(2)}$ production events are expected for 400 GeV $\leq 1/R \leq$ 
2000 GeV with an integrated luminosity of 100 fb$^{-1}$. We have also discussed the 
significance of various production processes and found that the KK number 
violating processes give leading contributions to the $\gamma^{(2)}$ and $Z^{(2)}$ 
productions for $1/R$ $\gtrsim$ 800 GeV.
%%%
In particular, our original processes (for example, $pp \rightarrow q^{(2)} 
q^{(0)}$, $pp \rightarrow q^{(2)} \bar q^{(0)}$, and so on) have provided large 
contributions to the $\gamma^{(2)}$ and $Z^{(2)}$ productions, and hence 
they have notable significance for the discrimination.
%%%

As a result of the calculations, we have shown the expected production number
of dileptons from the $\gamma^{(2)}$ and $Z^{(2)}$ decays (Table \ref{dilepton}). 
We have found that for $1/R \lesssim 1600$ GeV there is a chance to discriminate 
models by using the dilepton signals with 100 fb$^{-1}$ integrated luminosity.
%%%
For the discrimination, it is crucial whether the dilepton signals can be observed or 
not. We need further study to discuss the feasibility of the discovery of the MUED 
model by estimating the background from the SM. 
%%%

In addition, as discussed in Ref. \cite{Park:2009cs}, when $1/R$ is not so large, the 
bump-hunting for $g^{(2)}$ in the dijet invariant mass distribution might be one of the 
useful ideas for the complementary check of the discrimination by the dilepton signals\footnote{
Note that the simulation result of the $g^{(2)}$ invariant mass distribution of 
the dijet signal shown in Ref. \cite{Park:2009cs} cannot apply for the MUED model, because 
the coefficient of the KK number violating operator in their framework is different from the 
MUED model.}. However, since jet analysis is accompanied with SM QCD background, precise 
simulation is necessary for the bump-hunting for $g^{(2)}$.
%%%
We leave those for future work \cite{2009}.

%%%%%%%%%%%%%%%%%%%%%%%%%%%%%%%%%%%%%%%%%%%%%%%%%%%%%%%%%%%%%
\section*{Acknowledgments} %%%%%%%%%%%%%%%%%%%%%%%%%%%%%%%%%%%%%%%%%%%%%%
%%%%%%%%%%%%%%%%%%%%%%%%%%%%%%%%%%%%%%%%%%%%%%%%%%%%%%%%%%%%%
 The work of J. S. was supported in part by the Grant-in-Aid for the Ministry of Education, Culture, Sports, Science, 
and Technology, Government of Japan (No. 20025001, No. 20039001, and No. 20540251). The work of M. Y. was supported 
in part by the Grant-in-Aid for the Ministry of Education, Culture, Sports, Science, and Technology, Government of 
Japan (No. 20007555).

%%%%%%%%%%%%%%%%%%%%%%%%%%%%%%%%%%%%%%%%%%%%%%%%%%
\section*{Appendix : Decay rates of second KK particles} %%%%%%%%%%%%%%%%%%%%%%%
%%%%%%%%%%%%%%%%%%%%%%%%%%%%%%%%%%%%%%%%%%%%%%%%%%

In this Appendix we show analytic expressions for decay rates of second KK particles. The branching ratios of them (Fig. 9 - 12, 15, 
and 16) are obtained by using the following expressions. The decay rate of the second KK gauge boson $V^{(2)}$ is given by \\[0mm]
\begin{equation}
 \begin{split}
   &\Gamma_{ V^{(2)} \rightarrow f^{(2)} \bar f^{(0)}}
   = \frac{ C_{220} }{ 96 \pi } m_{V2}
   \sqrt{ \left(1 - \frac{m_{f2}^2 + m_{f0}^2}{m_{V2}^2} \right)^2 - 4\frac{m_{f2}^2 m_{f0}^2}{m_{V2}^4} ~}  \\
   & \times 
   \Biggl[~
      1 - \frac{m_{f2}^2 + m_{f0}^2}{m_{V2}^2} 
      + \left(1 + \frac{m_{f2}^2 - m_{f0}^2}{m_{V2}^2} \right) \left(1 - \frac{m_{f2}^2 - m_{f0}^2}{m_{V2}^2} \right) 
      + \alpha \frac{m_{f2} m_{f0}}{m_{V2}^2} ~
   \Biggr] ~,
 \end{split}
\end{equation} 

\begin{equation}
 \begin{split}
   \Gamma_{ V^{(2)} \rightarrow f^{(1)} \bar f^{(1)} }
   = \frac{ C_{211} }{ 48 \pi }  m_{V2}  
   \left[~ 1 - \frac{ 4 m_{f1}^2 }{ m_{V2}^2 } ~\right]^{3/2} ~,
 \end{split}
\end{equation} 

\begin{equation}
 \begin{split}
   \Gamma_{ V^{(2)} \rightarrow f^{(0)} \bar f^{(0)} } 
   = \frac{ m_{V2} }{ 24576 \pi }  
   \sqrt{ 1 - 4 \frac{m_{f0}^2}{m_{V2}^2}  }  ~
   \Biggl[~ 
      C_{200} (1 + \frac{ 2 m_{f0}^2 }{ m_{V2}^2 }) + C'_{200} (1 - \frac{ 4 m_{f0}^2 }{ m_{V2}^2 })   ~
   \Biggr] ~,
 \end{split} \label{V2f0fbar0}
\end{equation} 

\begin{equation}
 \begin{split}
   & \Gamma_{ Z^{(2)} \rightarrow h^{(1)} A^{(1)} } = \Gamma_{ Z^{(2)} \rightarrow H^{+(1)} H^{-(1)} }   \\[1mm]
   &  = \frac{ g_2^2 }{ 384 \pi }  m_{Z2} 
      \sqrt{  \left(1 - \frac{m_{H1}^2 + m_{H2}^2}{m_{Z2}^2} \right)^2 - 4\frac{m_{H1}^2 m_{H2}^2}{m_{Z2}^4} ~  }  \\
   & \times 
      \Biggl[~ 
         2 \left( 1 - \frac{ m_{H1}^2 + m_{H2}^2 }{m_{Z2}^2} \right) 
         - \left( 1 + \frac{ m_{H1}^2 - m_{H2}^2 }{m_{Z2}^2} \right) \left( 1 - \frac{ m_{H1}^2 - m_{H2}^2 }{m_{Z2}^2} \right)  ~
      \Biggr] ~.
 \end{split}
\end{equation}  \\[2mm]
Here $m_{V2}$, $m_{f2}$, $m_{f1}$, and $m_{f0}$ are the masses of $V^{(2)}$, the second 
KK fermion $f^{(2)}$, the first KK fermion $f^{(1)}$, and the SM fermion $f^{(0)}$. 
%%%
For $V^{(2)} = g^{(2)}$, $C_{220} = C_{211} = g_s^2$. For $V^{(2)} = Z^{(2)}$, 
$C_{200} = C_{211} = g_2^2/2$. $C_{200}$, and $C'_{200}$ are listed in Table \ref{tab:c}. 
%%%
The coefficient $\alpha$ depends on the mixing between the mass eigenstate and the interaction 
eigenstate of the KK fermions. The mixing is negligible except for the KK top quark, and hence 
$\alpha$ is as follows : %\\[0mm]
\begin{equation}
 \begin{split}
   & \alpha =  \frac{12 m_t}{ \sqrt{(\tilde m_{T^{(2)}} + \tilde m_{t^{(2)}})^2  +  4m_t^2 ~} }  
      ~~~~ \text{for}~ f^{(2)} = T^{(2)}   \\[1mm]
   & \alpha = \frac{- 12 m_t}{ \sqrt{(\tilde m_{T^{(2)}} + \tilde m_{t^{(2)}})^2  +  4m_t^2 ~} }    
      ~~~~ \text{for}~ f^{(2)} = t^{(2)}   \\[1mm]
   & \alpha = 0   ~~~~ \text{for}~ f^{(2)} = \text{other second KK fermions}  
 \end{split}
\end{equation}
Here $m_t$ stands for the SM top quark mass, and $\tilde m_{T^{(2)}}$ and $\tilde m_{t^{(2)}}$ are given by
\begin{equation}
 \begin{split}
    \tilde m_{T^{(2)}} &= \frac{2}{R} + \frac{2}{R} 
    \biggl[ 3 \frac{g_s^2}{16 \pi^2} + \frac{g'^2}{16 \pi^2} \biggr] \text{ln} \frac{\Lambda^2}{\mu^2}
    +  \frac{2}{R} \biggl[  - \frac{3}{2} \frac{y_t^2}{16 \pi^2}  \biggr]  \text{ln} \frac{\Lambda^2}{\mu^2} \\
    \tilde m_{t^{(2)}} &= \frac{2}{R} + \frac{2}{R} 
    \biggl[ 3 \frac{g_s^2}{16 \pi^2} + \frac{27}{16} \frac{g_2^2}{16 \pi^2} + 
    \frac{1}{16} \frac{g'^2}{16 \pi^2} \biggr] \text{ln} \frac{\Lambda^2}{\mu^2}
    +  \frac{2}{R} \biggl[  - \frac{3}{4} \frac{y_t^2}{16 \pi^2}  \biggr]  \text{ln} \frac{\Lambda^2}{\mu^2} ~.
 \end{split}
\end{equation}

\begin{table} [!ht]
\begin{center}
\begin{tabular}{|l||l|l|l|}
\hline
~~~~process  & ~~~~~~~~~~~~~~$C_{200}$~ & ~~~~~~~~~~~~~~$C'_{200}$~  \\
\hline\hline
\rule[-13pt]{0pt}{31pt}
$\gamma^{(2)} \rightarrow e^{(0)} \bar e^{(0)}$  
                 & $ (g'^2/2) \bigl[ 85 c' + 27 c_2 \bigr]^2 $
                 & $ (g'^2/2) \bigl[ (193/3) c' - 27 c_2 \bigr]^2 $  \\
\hline      
\rule[-13pt]{0pt}{31pt}
$\gamma^{(2)} \rightarrow \nu^{(0)} \bar \nu^{(0)}$  
                 & $ (g'^2/2) \bigl[ (31/3) c' + 27 c_2 \bigr]^2 $
                 & $ (g'^2/2) \bigl[ (31/3) c' + 27 c_2 \bigr]^2 $  \\
\hline    
\rule[-13pt]{0pt}{31pt}
$\gamma^{(2)} \rightarrow u^{(0)} \bar u^{(0)}$ 
                 & \parbox{150pt}{$ (g'^2/6)$ \\$ \times \bigl[ 215 c' + 81 c_2 + 720 c_s \bigr]^2 $}
                 & \parbox{150pt}{$ (g'^2/2)$ \\$ \times \bigl[ 67 c' - 27 c_2 + 144 c_s \bigr]^2 $}  \\
\hline      
\rule[-13pt]{0pt}{31pt}
$\gamma^{(2)} \rightarrow d^{(0)} \bar d^{(0)}$  
                 & \parbox{150pt}{$ (g'^2/6) $ \\$  \times\bigl[ 25 c' - 81 c_2 + 144 c_s \bigr]^2 $}
                 & \parbox{150pt}{$ (g'^2/2) $ \\$  \times \bigl[ 13 c' + 27 c_2 + 144 c_s \bigr]^2 $}  \\
\hline   
\rule[-13pt]{0pt}{31pt}
$\gamma^{(2)} \rightarrow t^{(0)} \bar t^{(0)}$  
                 & \parbox{150pt}{$ (g'^2/6) \times$ \\$ \bigl[ 215 c' + 81 c_2 + 720 c_s + 252 c_t \bigr]^2 $}
                 & \parbox{150pt}{$ (g'^2/2) \times$ \\$ \bigl[ 67 c' - 27 c_2 + 144 c_s + 76 c_t \bigr]^2 $}  \\
\hline  
\rule[-13pt]{0pt}{31pt}
$\gamma^{(2)} \rightarrow b^{(0)} \bar b^{(0)}$  
                 & \parbox{150pt}{$ (g'^2/6) \times$ \\$  \bigl[ 25 c' - 81 c_2 + 144 c_s - 12 c_t \bigr]^2 $}
                 & \parbox{150pt}{$ (g'^2/2)\times $ \\$  \bigl[ 13 c' + 27 c_2 + 144 c_s + 4 c_t \bigr]^2 $}  \\
\hline
\rule[-13pt]{0pt}{31pt}
$Z^{(2)} \rightarrow e^{(0)} \bar e^{(0)}$  
                 & $ (g_2^2/18) \bigl[ 27 c' - 95 c_2 \bigr]^2 $
                 & $ (g_2^2/18) \bigl[ 27 c' - 95 c_2 \bigr]^2 $  \\
\hline
\rule[-13pt]{0pt}{31pt}
$Z^{(2)} \rightarrow \nu^{(0)} \bar \nu^{(0)}$  
                 & $ (g_2^2/18) \bigl[ 27 c' - 95 c_2 \bigr]^2 $
                 & $ (g_2^2/18) \bigl[ 27 c' - 95 c_2 \bigr]^2 $  \\
\hline
\rule[-13pt]{0pt}{31pt}
$Z^{(2)} \rightarrow u^{(0)} \bar u^{(0)}$  
                 & $ (g_2^2/6) \bigl[ 3 c' - 95 c_2 + 144 c_s \bigr]^2 $
                 & $ (g_2^2/6) \bigl[ 3 c' - 95 c_2 + 144 c_s \bigr]^2 $  \\
\hline
\rule[-13pt]{0pt}{31pt}
$Z^{(2)} \rightarrow d^{(0)} \bar d^{(0)}$  
                 & $ (g_2^2/6) \bigl[ 3 c' - 95 c_2 + 144 c_s \bigr]^2 $
                 & $ (g_2^2/6) \bigl[ 3 c' - 95 c_2 + 144 c_s \bigr]^2 $  \\
\hline
\rule[-13pt]{0pt}{31pt}
$Z^{(2)} \rightarrow t^{(0)} \bar t^{(0)}$  
                 & \parbox{150pt}{$ (g_2^2/6)  \times $ \\ $ \bigl[ 3 c' - 95 c_2 + 144 c_s + 12 c_t \bigr]^2 $}
                 & \parbox{150pt}{$ (g_2^2/6) \times $ \\ $ \bigl[ 3 c' - 95 c_2 + 144 c_s + 12 c_t \bigr]^2 $}  \\
\hline
\rule[-13pt]{0pt}{31pt}
$Z^{(2)} \rightarrow b^{(0)} \bar b^{(0)}$  
                 & \parbox{150pt}{$ (g_2^2/6) \times $ \\$ \bigl[ 3 c' - 95 c_2 + 144 c_s + 12 c_t \bigr]^2 $}
                 & \parbox{150pt}{$ (g_2^2/6) \times $ \\ $ \bigl[ 3 c' - 95 c_2 + 144 c_s + 12 c_t \bigr]^2 $}  \\
\hline
\rule[-13pt]{0pt}{31pt}
$g^{(2)} \rightarrow u^{(0)} \bar u^{(0)}$   %(or $c^{(0)} \bar c^{(0)}$) 
                 & $ g_s^2 \bigl[ 17 c' + 27 c_2 - 88 c_s \bigr]^2 $ 
                 & $ g_s^2 \bigl[ 15 c' - 27 c_2  \bigr]^2 $   \\
\hline
\rule[-13pt]{0pt}{31pt}
$g^{(2)} \rightarrow d^{(0)} \bar d^{(0)}$   %(or $s^{(0)} \bar s^{(0)}$)  
                 & $ g_s^2 \bigl[ 5 c' + 27 c_2 - 88 c_s \bigr]^2 $ 
                 & $ g_s^2 \bigl[ 3 c' - 27 c_2  \bigr]^2 $  \\
\hline
\rule[-13pt]{0pt}{31pt}
$g^{(2)} \rightarrow t^{(0)} \bar t^{(0)}$  
                 & $ g_s^2 \bigl[ 17 c' + 27 c_2 - 88 c_s + 8 c_t \bigr]^2 $
                 & $ g_s^2 \bigl[ 15 c' - 27 c_2  \bigr]^2 $  \\
\hline
\rule[-13pt]{0pt}{31pt}
$g^{(2)} \rightarrow b^{(0)} \bar b^{(0)}$  
                 & $ g_s^2 \bigl[ 5 c' + 27 c_2 - 88 c_s + 4 c_t \bigr]^2 $
                 & $ g_s^2 \bigl[ 3 c' - 27 c_2 -4 c_t \bigr]^2 $  \\
\hline
\end{tabular}
\caption{{\small{Coefficients $C_{200}$ and $C'_{200}$ in Eq. (\ref{V2f0fbar0}}). Here $c'$, 
                      $c_2$, $c_s$, and $c_t$ are given in Eq. (\ref{c_i}).
           }}
\label{tab:c}
\end{center}
\end{table}

The decay rate of the first and second generation second KK quarks are given by   \\[1mm]
\begin{equation}
 \begin{split}
   &\Gamma_{ q^{(2)} \rightarrow V^{(2)} q^{(0)} } 
     =
     \frac{K}{16 \pi} m_{q2} 
     \sqrt{ \left( 1 - \frac{m_{V2}^2 + m_{q0}^2}{m_{q2}^2} \right)^2 -4 \frac{m_{V2}^2 m_{q0}^2}{m_{q2}^4} ~}     \\
   &\times 
     \Biggl[
        1 - \frac{m_{V2}^2 - m_{q0}^2}{m_{q2}^2} 
        +
        \Biggl( 1 + \frac{m_{q2}^2 - m_{q0}^2}{m_{V2}^2} \Biggr)   
        \Biggl( 1 - \frac{m_{V2}^2 + m_{q0}^2}{m_{q2}^2} \Biggr) 
     \Biggr]   ~,
 \end{split}
\end{equation} 

\begin{equation}
 \begin{split}
   &\Gamma_{ q^{(2)} \rightarrow V^{(1)} q^{(1)} } 
     = 
     \frac{K}{16 \pi} m_{q2}  
     \sqrt{ \left( 1 - \frac{m_{V1}^2 + m_{q1}^2}{m_{q2}^2} \right)^2 -4 \frac{m_{V1}^2 m_{q1}^2}{m_{q2}^4} ~}     \\
   &\times 
     \Biggl[
        1 + \frac{m_{q1}^2 - m_{V1}^2}{m_{q2}^2} -6 \frac{m_{q1}}{m_{q2} }
        +
        \Biggl( 1 + \frac{m_{q2}^2 - m_{q1}^2}{m_{V2}^2} \Biggr)   
        \Biggl( 1 - \frac{m_{V2}^2 + m_{q1}^2}{m_{q2}^2} \Biggr) 
     \Biggr]  ~.
 \end{split}
\end{equation}   \\[1mm]
Here $m_{q2}$, $m_{q1}$, and $m_{q0}$ are the masses of $q^{(2)}$,
$q^{(1)}$, and $q^{(0)}$. $K = g'^2/72$ for ($q,~ V$) = ($Q,~ \gamma$),
$K = g_2^2/4$ for ($q,~ V$) = ($Q,~ W^\pm$), $K = g_2^2/8$ for ($q,~ V$) =
($Q,~ Z$), $K = 2 g'^2/9$ for ($q,~ V$) = ($U,~ \gamma$), and $K =
g'^2/18$ for ($q,~ V$) = ($D,~ \gamma$).

\end{document}